\documentclass[12pt]{amsart}
\usepackage{amssymb}
\usepackage{verbatim}
\usepackage[usenames]{color}
\usepackage{hyperref}

\newtheorem{thm}{Theorem}
\newtheorem{prop}[thm]{Proposition}
\newtheorem{la}[thm]{Lemma}
\newtheorem{cor}[thm]{Corollary}

\theoremstyle{definition}
\newtheorem{df}[thm]{Definition}

\newtheorem{notat}[thm]{Notation}

\theoremstyle{remark}

\newtheorem{rmk}[thm]{Remark}

\newenvironment{ls}{\begin{itemize}}{\end{itemize}}
\newenvironment{lsnum}{\begin{enumerate}}{\end{enumerate}}
\newenvironment{pf}{\begin{proof}}{\end{proof}}

\newcommand{\scr}[1]{\ensuremath{\mathcal {#1}}}

\newcommand{\bbb}[1]{\ensuremath{\mathbb {#1}}}

\renewcommand{\phi}{\varphi}
\newcommand{\eps}{\varepsilon}
\newcommand{\sq}[1]{\ensuremath{\langle#1\rangle}}

\newcommand{\notarrow}{\kern .42em\not\kern -.42em\longrightarrow}

\newcommand{\ket}[1]{\ensuremath{|#1\rangle}}
\newcommand{\bra}[1]{\ensuremath{\langle#1|}}
\newcommand{\tr}[1]{\ensuremath{\text{Tr}(#1)}}
\newcommand{\nm}[1]{\ensuremath{\Vert #1 \Vert}}
\newcommand{\cv}[1]{\ensuremath{\text{Conv}(#1)}}
\newcommand{\af}[1]{\ensuremath{\text{Aff}(#1)}}

\newcommand{\noprint}[1]{\relax}

\title[Hidden Variables:
Value and Expectation No-Go Theorems]%
{On Hidden Variables:\\
Value and Expectation No-Go Theorems}
\author{Andreas Blass}
\address{Mathematics Department\\
University of Michigan\\
Ann Arbor, MI 48109--1043, U.S.A.}
\email{ablass@umich.edu}
\thanks{Part of the first author's work was done as a visiting
  researcher at Microsoft Research; another part was done as a
  visiting fellow at the Isaac Newton Institute for Mathematical
  Sciences.}
\author{Yuri Gurevich}
\address{Microsoft Research, One Microsoft Way, Redmond, WA 98052,
  U.S.A.}
\email{gurevich@microsoft.com}

\begin{document}

\begin{abstract}
  No-go theorems assert that hidden-variable theories, subject to
  appropriate hypotheses, cannot reproduce the predictions of quantum
  theory. We examine two species of such theorems, value no-go
  theorems and expectation no-go theorems. The former assert that
  hidden-variables cannot match the predictions of quantum theory
  about the possible values resulting from measurements; the latter
  assert that hidden-variables cannot match the predictions of quantum
  theory about the expectation values of measurements. We sharpen the
  known results of both species, which allows us to clarify the
  similarities and differences between the two species.  We also
  repair some flaws in existing definitions and proofs.
\end{abstract}

\maketitle

\section{Introduction}          \label{intro}

This paper is about ``no-go'' theorems asserting the impossibility of
schemes for explaining the probabilistic aspects of quantum mechanics
in terms of ordinary, classical probability.  Such schemes are often
called hidden-variable theories.  They postulate that a quantum state,
even if it is a pure state and thus contains as much information as
quantum mechanics permits, actually describes an ensemble of systems
with different values for some additional, hidden properties that are
not taken into account in quantum mechanics.  The ensemble given by a
quantum state is thus composed of sub-ensembles, each having specific
values for the hidden variables.  The idea is that, once
the values of these hidden variables are specified, all the properties
of the system become determinate (or at least more determinate than
quantum mechanics says).  Thus the randomness in quantum predictions
results (entirely or at least partially) from the randomness involved
in choosing a particular element, with particular values of the hidden
variables, from the ensemble that a quantum state describes.

No-go theorems for hidden-variable interpretations of quantum
mechanics assert that, under reasonable assumptions, a hidden-variable
interpretation cannot reproduce the predictions of quantum mechanics.
There are many no-go theorems in the literature. Although they all
share the basic idea, ``hidden-variable theories cannot succeed,''
they differ from one another in the particular description of what a
hidden-variable theory is and what is meant by succeeding.  A typical
no-go theorem can be formulated in terms of a hypothesis saying what a
hidden-variable theory should look like and a conclusion saying that
certain predictions of quantum mechanics can never result from such a
theory.  In this paper, we examine two species of such theorems,
\emph{value} no-go theorems and \emph{expectation} no-go theorems. We
sharpen the results of both species, which allows us to clarify both
the similarities and the differences between the two species.

The value approach originated in the work of Bell \cite{bell64,bell66}
and of Kochen and Specker \cite{ks} in the 1960's.  A very readable
overview of this work, with some simplifications and historical
information, is given by Mermin \cite{mermin}.
Value no-go theorems establish that, under suitable hypotheses,
hidden-variable theories cannot reproduce the predictions of quantum
mechanics concerning the possible results of measurements.  There is
no need to consider the probabilities of possible results or the
expectation values of measurements; the measured values alone provide
a discrepancy between hidden-variable theories and quantum theory.
The hypotheses that are used to deduce these theorems concern the
measurements of observables in quantum states.

The expectation approach was developed in the last decade by Spekkens
\cite{spekkens} and by Ferrie, Emerson, and Morris
\cite{ferrie2008,ferrie2009,ferrie2010}, with \cite{ferrie2010} giving
the sharpest result.  In this approach, the discrepancy between
hidden-variable theories and quantum mechanics appears in the
predictions of the expected values of measurements.  There is no need
to consider the actual values obtained by measurements or the
probability distributions over these values.
The hypotheses that are used to deduce these results concern the
measurement of effects, i.e.\ the elements of positive operator-valued
measurements (POVMs).  Effects are represented by Hermitian operators
with spectrum on the real interval $[0,1]$. They are regarded as
representing yes-or-no questions, the probability of ``yes'' for
effect $E$ in state \ket\psi\ being $\bra\psi E\ket\psi$.

Although both approaches involve measurements associated to Hermitian
operators, they are different sorts of measurements.  In the value
approach, Hermitian operators serve as observables, and measuring one
of them produces a number in its spectrum.  In the expectation
approach, certain Hermitian operators serve as effects, and measuring
one of them produces ``yes'' or ``no'', i.e., 1 or 0, even if the
spectrum contains --- or even consists entirely of --- other points.
The only Hermitian operators for which these two uses coincide are the
projections, the operators whose spectrum is included in $\{0,1\}$. We
sharpen the results of both approaches so that only projection
measurements are used.

The present work started with repairing various flaws in the
literature on expectation no-go theorems.  Although the papers purport
to specify the exact assumptions needed to obtain their no-go results,
some of them bring in, afterward, an additional assumption of
convex-linearity; another erroneously claims that this assumption
follows from the others.  In addition, the assumptions are sometimes
ambiguous, and one of the papers relies on an erroneous result of
Bugajski \cite{bugajski}, which needs some additional hypotheses to
become correct. We explain the flaws that we found and how to
circumvent them, and we strengthen the Ferrie, Morris, and Emerson
result in \cite{ferrie2010} by substantially weakening the
hypotheses. We do not need arbitrary effects, or even arbitrary sharp
effects, but only rank-1 projections. Accordingly, we need
convex-linearity only for the hidden-variable picture of states, not
for that of effects.

\begin{thm}     \label{exp-thm-intro}
  For no quantum system are there a measurable space $\Lambda$, a
  convex-linear map $T$ from density matrices $\rho$ to probability
  measures on $\lambda$, and a map $S$ from rank-1 projections $E$ in
  the Hilbert space of the system into measurable functions from
  $\Lambda$ to $[0,1]$, such that $ \tr{\rho E}=\int_\Lambda
  S(E)\,dT(\rho) $ for all $\rho$ and $E$.
\end{thm}

Some of the literature on expectation no-go theorems emphasizes a
symmetry between states and effects.  We explain why such symmetry is
to be expected only when the Hilbert space of states is
finite-dimensional and the space of possible values for the
hidden-variables is not merely finite-dimensional but finite.

We formulate the value no-go theorems in terms of the maps, from
observables to real numbers, that a hidden-variable theory would
assign to individual systems.  We define a \emph{value map} $v$ for a
set $\scr O$ of Hermitian operators on Hilbert space $\scr H$ to be a
function that assigns to each operator $A\in\scr O$ a number $v(A)$ in the
spectrum of $A$ in such a way that, for any pairwise commuting
operators $A_1,\dots,A_n\in\scr O$, the tuple $(v(A_1),\dots,v(A_n))$
belongs to the joint spectrum of the tuple $(A_1,\dots,A_n)$. (The
notion of joint spectra is uncommon in the quantum literature, so we
explain it and its relevant properties.)  Our value no-go theorem is
close to one of Bell's results, as interpreted by Mermin
\cite{mermin}.

\begin{thm}     \label{val-thm-intro}
Suppose that $\scr H$ is a Hilbert space of dimension $\ge3$.
\begin{enumerate}
\item There is a finite set of projections for which no value map exists.
\item If $\dim(\scr H)<\infty$ then there is a finite set of rank-1
  projections for which no value map exists.
\end{enumerate}
\end{thm}

The desired finite sets of projections are constructed explicitly in
the proof. The condition $\dim(\scr H)\ge3$ is necessary. In the case of
$\dim(\scr H)=2$ there are counterexamples \cite{bell66,mermin} that
produce not only correct values but also correct probabilities for
pure states; we slightly simplify the verification of that. These
counterexamples do not violate Theorem~\ref{exp-thm-intro} merely
because they apply only to pure states and do not admit convex
linearity.  The condition $\dim(\scr H)<\infty$ in (2) is also
necessary. If $\dim(\scr H)=\infty$ then the zero function is a value map
for the set of all finite-rank projections.

Note that there is no implication in either direction between our two
theorems. One says that a hidden-variable theory cannot predict the
correct values for measured quantities (though it might predict
correct expectations) while the other says that a hidden-variable
theory cannot predict the correct expectations (though it might
predict the correct values, with incorrect probabilities). Thus, there
are two separate reasons why hidden theories must fail.

We postpone to future work a similar study of no-go theorems for local
hidden-variable theories.  In these theories, a certain amount of
contextuality is allowed, which means that the measured value of an
observable can depend on which other, commuting observables are
measured along with it, but only if those other observables are local
in a suitable sense.  There are value no-go theorems for such theories
\cite{bell64,mermin}, but they rely on a stronger notion of value
map. Consider, for example, two observables that do not commute and
therefore cannot, in general, be simultaneously measured. They might
nevertheless share a common eigenvector \ket\psi\ and would then have
simultaneous definite values when the state of the system is \ket\psi.
In this case, a hidden-variable theory should provide a value map that
assigns appropriately correlated values for these two observables.

This paper is organized as follows.  In Section~\ref{variants}, we
describe in detail the ingredients of various hidden-variable
theories.  Section~\ref{exp} is devoted to expectation no-go theorems.
We begin by describing the work of Spekkens \cite{spekkens} and of
Ferrie, Emerson, and Morris \cite{ferrie2008,ferrie2009,ferrie2010},
pointing out the flaws that we found and suggesting how to circumvent
them.  At the end of the section, we prove our expectation no-go
result, Theorem~\ref{exp-thm-intro}, which strengthens the result of
Ferrie, Morris, and Emerson in \cite{ferrie2010}.  Section~\ref{val}
is devoted to value no-go theorems, ending with the proof of
Theorem~\ref{val-thm-intro}.

Section~\ref{reduce} is devoted to giving a mathematical basis for the
intuitive idea that a hidden-variable theory for one Hilbert space
should specialize to a hidden-variable theory for any closed subspace,
because the latter space just represents a subset of the states of the
former.  Thus a no-go theorem for the subspace should imply a no-go
theorem for the larger space.  We prove theorems that support this
intuition in several cases.

Section~\ref{example} examines an example, due to Bell \cite{bell66}
and described by Mermin \cite{mermin}, of a hidden-variable theory for
pure states in the case of a two-dimensional Hilbert space.  The
example shows that the assumption of dimension at least 3 cannot be
omitted from Theorem~\ref{val-thm-intro}.  Our expectation no-go
result, Theorem~\ref{exp-thm-intro}, applies in all dimensions from 2
up, and we point out why the example does not contradict the theorem.

The paper has two appendices.  The first discusses the notion of
convex-linearity, which played a role in some of the flaws we found in
the literature.  The second presents a no-go theorem adapted to the
original framework described by Spekkens \cite{spekkens}, minimally
modified to remove unintended aspects and ambiguities.

\section{Hidden-Variable Theories}      \label{variants}

In this section, we describe some of the differences between various
approaches to hidden variables.  These differences include what sorts
of quantum states are considered, what sorts of measurements are
considered, and which predictions of quantum mechanics should be
matched by the hidden-variable theory.

\subsection{States}

Most hidden-variable theories begin with states in the usual sense of
quantum mechanics and seek to make their properties more determinate
by adjoining hidden variables.  In some cases, however, they begin
with a more primitive notion, that of a \emph{preparation}, a way of
producing systems in a specific quantum state. Different preparations
might produce the same state.  In \cite{spekkens}, Spekkens works with
a notion of \emph{ontological model} of quantum theory,
in which distribution functions (describing how a quantum ensemble is
composed of more determinate sub-ensembles) are assigned to
preparation procedures.  He gives the name ``preparation
noncontextuality'' to the hypothesis that different preparations of
the same quantum state yield the same distribution function, i.e.,
that the distribution function is determined by the quantum state.
This hypothesis is in force for most of \cite{spekkens}, but it is
pointed out explicitly as a hypothesis that could, in principle, be
questioned.

Other hidden-variable theories begin with quantum states rather than
with preparations, so that preparation noncontextuality is built into
the foundational framework of these theories.  They seek to analyze
quantum states as ensembles obtained by mixing sub-ensembles with more
determinate properties.  These sub-ensembles are viewed in different
ways by the various theories, but these viewpoints are ultimately
equivalent.  For example, Mermin \cite{mermin} talks about individual
systems in the quantum ensemble while von~Neumann \cite{neumann} talks
about dispersion-free sub-ensembles.  Other authors \cite{spekkens,
  ferrie2008, ferrie2009, ferrie2010}, do not refer to the
sub-ensembles explicitly but work with distribution functions over a
space whose points are best viewed as parametrizing such
sub-ensembles.

Even after one decides to work with quantum states, one still has a
choice whether to work only with pure states or to admit mixed states
as well.  At first sight, the difference between these two options
might seem unimportant.  After all, any mixed state is a weighted
average of pure states.  So, given interpretations of pure states as
ensembles, we can use weighted mixtures of these ensembles to
represent mixed states.  The situation is, however, more subtle.  A
single mixed state may be represented as a weighted average of pure
states in more than one way.  Can the associated weighted averages of
ensembles depend on which of these representations we use? In general,
the answer is yes, and then we do not obtain a single, well-defined
ensemble to represent this mixed state.  Well-definedness of the
ensemble representations of mixed states is not automatic but rather
imposes a non-trivial consistency requirement on the representations
of the pure states.  In Section~\ref{example}, we shall describe an
example, essentially due to Bell, of a hidden-variable representation
of pure states (for a 2-dimensional Hilbert space) that cannot be
extended to mixed states while respecting weighted averages.

To summarize this situation, we list four approaches to the issue of
what states (or preparations) should be given a hidden-variable
interpretation.  (We use ``mixed'' here to mean ``possibly mixed'';
pure states are included among the mixed ones.)
\begin{lsnum}
  \item Pure states, with no consistency requirement on the
    representation.
\item Pure states, subject to the consistency requirement allowing a
  well-defined extension to mixed states, by respecting weighted
  averages.
\item Mixed states, with no consistency requirement.
\item Mixed states, subject to the requirement of respecting weighted
  averages.
\end{lsnum}
In items~2 and 4, ``respecting weighted averages'' means that the
collection of sub-ensembles associated to a weighted mixture of some
given states is the corresponding weighted average of the
sub-ensembles for the given states.  In item~2, respect for weighted
averages serves as a method for extending the hidden-variable
interpretation from pure to mixed states.  In item~4, respect for
weighted averages is a requirement imposed on the assumed
interpretation of mixed states.  These two items are equivalent, in
the sense that the mixed-state interpretations considered in item~4
are exactly the (unique) extensions to mixed states of the pure-state
interpretations considered in item~2.

The other two items in the list, items~1 and 3, are more liberal
because they do not require any respect for weighted averages.

The preceding list of four (or three in view of the equivalence
between items~2 and 4) approaches could be doubled by including
analogous versions with preparations in place of states.

\begin{notat} \label{testing} The concept of respecting weighted
  averages has several names in the literature.  The formal definition
  of the concept, namely that the function $f$ in question satisfies
  $f(ax+by)=af(x)+bf(y)$ whenever $a$ and $b$ are nonnegative real
  numbers with sum 1, looks like the definition of linearity except
  that it applies only to the restricted options for $a$ and $b$ that
  produce weighted averages, also known as convex combinations.
  Because of this, some authors, for example Spekkens \cite{spekkens},
  use the term \emph{convex linear}, and we shall follow this
  terminology.  Other authors (see, for example,
  \cite{ferrie2008,ferrie2009,ferrie2010}) prefer the shorter name
  \emph{affine}, though this would seem more natural for the related
  concept where $a$ or $b$ can be negative and the only constraint on
  them is $a+b=1$.  In Appendix~\ref{conv-lin}, we look more closely
  at the notion of convex-linearity.
\end{notat}

\subsection{Measurements}

We consider next the sorts of measurements that a hidden-variable
theory should explain.  In quantum mechanics, measurements are
ordinarily represented by certain Hermitian operators on the Hilbert
space of states of a system.  In this context, those operators are
usually called observables.

Before turning to the question of which operators should be treated in
a hidden-variable theory, we first address a prior issue, analogous to
the issue of state versus preparation in the previous subsection.  The
analogous issue here is measurement versus apparatus.  It is entirely
possible that different experimental arrangements measure the same
observable.  In such a situation, those arrangements should produce
the same results (the same statistical distribution of measured
values) for any particular quantum state, but it is not clear that
they should produce the same results on each of the sub-ensembles
considered in a hidden-variable theory.  Spekkens's ontological models
\cite{spekkens} assign measurement values not to observables but to
measurement procedures.  He introduces the name ``measurement
noncontextuality'' for the hypothesis that different measurement
procedures for the same observable result in the same outcomes.
(Actually, he deals only with measurements of effects; see below.)

When hidden-variable theories take observables to be the entities to
be measured in their sub-ensembles, either because of an explicit
assumption of measurement noncontextuality or because observables are
built into the foundation of the theory, there still remains a choice
as to which observables are to be considered and what is meant by
measuring them.

A traditional viewpoint is that observables are arbitrary%
\footnote{We ignore here the complications arising from superselection
  rules, which make some Hermitian operators unobservable.}%
Hermitian operators and that a measurement of such an operator in some
state produces a real number in the spectrum of the operator.  For
simplicity, we shall pretend for a while that our Hilbert spaces are
finite-dimensional, so that a measurement produces an eigenvalue of
the operator. We shall see in Section~\ref{reduce} that no-go theorems
for finite-dimensional Hilbert spaces typically imply the
corresponding theorems for infinite-dimensional spaces, so in these
cases our simplifying assumption does not really lose generality.
Quantum mechanics gives well-known formulas for the probabilities of
the various eigenvalues and therefore also for quantities like the
expectation of the measured values.

For a hidden-variable theory to successfully match the predictions of
quantum mechanics, one would reasonably require it to predict, in
particular, the possible values of any measurement (namely the
eigenvalues of the observable being measured) and their respective
probabilities.  It turns out, somewhat surprisingly, that several
no-go theorems work under considerably weaker demands on what the
hidden-variable theory must accomplish.  Specifically, some theorems
show that a hidden-variable theory cannot even predict the correct
values for all observables, even if one doesn't care about
probabilities or even the expectation values.  Other theorems show
that a hidden-variable theory cannot even predict the correct
expectations, even if one doesn't care about the particular values or
probabilities.  For brevity, we shall refer to theorems of these two
sorts as ``value no-go'' and ``expectation no-go'' theorems,
respectively.

Another common view of measurements in quantum mechanics is based not
on observables but on particular Hermitian operators called
\emph{effects} and on certain sets of effects called \emph{positive
  operator-valued measures} (POVMs).  An effect is a Hermitian
operator $E$ whose spectrum lies in the interval $[0,1]$ of the real
line.  Among the effects are the \emph{sharp effects}, those whose
spectrum is included in the two-element set $\{0,1\}$.  The sharp
effects are simply the projection operators from the Hilbert space
onto its closed subspaces.  Arbitrary effects are weighted averages of
sharp ones.  A POVM is a set of effects whose sum is the identity
operator $I$.  Notice that every effect $E$ is a member of at least
one POVM, namely $\{E,I-E\}$; unless $E=I$, it is also a member of
numerous other POVMs.

A POVM $\{E_k:k\in S\}$, where $S$ is some index set (usually finite),
is intended to model a measurement whose outcome is a member of $S$,
the probability of outcome $k$ for state \ket\psi\ being $\bra\psi
E_k\ket\psi$, or, in the case of mixed states with density matrix
$\rho$, \tr{E_k\rho}.  Measurement of an observable $A$ amounts to
measurement of the POVM consisting of the projections to the
eigenspaces of $A$.  Arbitrary POVMs are more general in two respects,
first that the effects in a POVM need not be sharp, and second that
these effects need not commute with one another.  Despite the
additional generality, it is known that general POVM measurements can
be reduced to measurements of observables in a larger Hilbert space,
one in which the original Hilbert space is isometrically embedded.
For details, see for example \cite[Section~5.1]{watrous} or
\cite{wiki-povm}.  Because actually measuring a general POVM can be a
complicated process, involving an enlargement of the original Hilbert
space, it is not clear that POVMs are so fundamental that a
hidden-variable theory should be required to produce correct
predictions for them.  In particular, it is not clear that enlargement
of the Hilbert space makes sense for the sub-ensembles considered by
such theories.  It is therefore preferable for no-go theorems to apply
even when the hidden-variable theory is required to work correctly
only for those POVMs whose measurement does not require enlarging the
Hilbert space.  Such POVMs include, in particular, those consisting of
mutually commuting, sharp effects.

It makes sense to speak of measuring a single effect $E$; this means
measuring the POVM $\{E,I-E\}$.  In other words, it is a yes-or-no
measurement, with ``yes'' corresponding to $E$ and ``no'' to $I-E$.
The probability of the answer ``yes'' when effect $E$ is measured in a
pure state $\ket\psi$ is $\bra\psi E\ket\psi$; for a mixed state with
density matrix $\rho$, it is the trace \tr{E\rho}.

When a hidden-variable theory uses POVMs and effects as the
measurements for which values are predicted, we encounter a third
notion of noncontextuality, in addition to the preparation
noncontextuality and measurement noncontextuality mentioned above.
The question here is whether the measurement of an effect $E$ depends
only on $E$ itself or on the entire POVM of which $E$ is a member.
For a quantum state, the probability of getting ``yes'' when measuring
$E$ depends only on $E$, but that does not necessarily imply that the
same situation obtains for all the sub-ensembles within that state.
The assertion that, even for the sub-ensembles, it is only $E$ that
matters, not the whole POVM, is the third sort of noncontextuality.

This issue also arises, as made very clear in \cite{mermin}, when
measurements are given by observables rather than effects and POVMs.
Noncontextuality in this context means that the result of measuring an
observable $A$ does not depend on what other observables might be
measured along with $A$.  (In this framework, those other observables
must commute with $A$ and with each other, for otherwise they could
not be measured simultaneously.  The framework does not envision
enlarging the Hilbert space.)

We shall use the word \emph{determinate} to refer to all sorts of
noncontextuality.  The intended meaning is that a hidden-variable
theory's analysis of some aspect of quantum theory --- such as states
or observables or effects --- should be completely determined by what
is explicitly mentioned, regardless of other aspects of the situation
--- preparations or apparatuses or other simultaneous measurements.

\begin{rmk}             \label{eff-op}
  Before leaving the discussion of measurements, we point out, to avoid
possible confusion, that, although an effect $E$ is, in particular, a
Hermitian operator and thus an observable, measuring it as an effect
is quite different from measuring it as an observable.  According to
quantum theory, the former always produces 1 (``yes'') or 0 (``no'');
the latter always produces one of the eigenvalues of $E$.  The two
sorts of measurement coincide only when $E$ is a sharp effect.
\end{rmk}

\section{Expectation No-Go Theorems}          \label{exp}

In this section, we discuss, clarify, and extend the work of Spekkens
\cite{spekkens} and of Ferrie et
al. \cite{ferrie2008,ferrie2009,ferrie2010} , which yields what we
called expectation no-go theorems above.  That is, under suitable
hypotheses, it is shown that hidden-variable theories cannot correctly
predict the expectation values of effects.  To describe and clarify
the contents of these papers, we begin with the earliest of them,
\cite{spekkens}, comment on various aspects in need of clarification,
and then indicate how these aspects are developed in the three papers
of Ferrie et al.

The papers under discussion differ somewhat in the hypotheses that
they explicitly assume, and they also differ in their names for the
theories that satisfy their hypotheses.  We shall use the generic name
``probability representations'' for these theories.  In the rest of
this section, we shall describe in considerable detail the variations
in content of these theories; see Remark~\ref{lang} for variations in
the terminology.

\subsection{Spekkens's no-go theorem}   \label{spek:subsec}

The following definition is essentially from \cite{spekkens}, but see
the commentary following the definition for more details.

\begin{df}              \label{v1}
  Given a Hilbert space \scr H, a \emph{probability representation
    (Spekkens version)} for quantum systems described by \scr H consists of
\begin{ls}
\item a measure space $\Lambda$,
\item for every density operator $\rho$ on \scr H, a nonnegative
  real-valued measurable function $\mu_\rho$ on $\Lambda$, normalized
  so that $\int_\Lambda\mu_\rho(\lambda)\,d\lambda=1$,
\item for every POVM $\{E_k\}$, a set $\{\xi_{E_k}\}$ of nonnegative
  real-valued measurable functions on $\Lambda$ that sum to the unit
  function on $\Lambda$, subject to the requirement that, if $E_k=0$,
  then the associated function $\xi_{E_k}$ is identically zero,
\end{ls}
such that for all density operators $\rho$ and all POVM elements
$E_k$, we have $\text{Tr}(\rho E_k)=\int_\Lambda
d\lambda\,\mu_\rho(\lambda)\xi_{E_k}(\lambda)$.
\end{df}

The intention behind this definition is that each point
$\lambda\in\Lambda$ represents a particular sub-ensemble as provided
by the hidden-variable theory.  A quantum state $\rho$ represents an
ensemble composed of various of these sub-ensembles, mixed together
according to the probability measure $\mu_\rho(\lambda)\,d\lambda$.
When an effect $E$ is measured on a system from the sub-ensemble
$\lambda$, the probability of getting a ``yes'' answer is
$\xi_E(\lambda)$.  Note that even a sharp effect can have, in a
sub-ensemble $\lambda$, a non-trivial probability of producing ``yes'';
the probability need not be 0 or 1.  This is discussed in detail in
the early part of \cite{spekkens}.

Note that the last part of Definition~\ref{v1} requires the
expectation value for an effect $E$ in a state $\rho$, as computed by
quantum mechanics, namely \tr{\rho E}, to agree with the prediction of
the hidden-variable theory, the weighted average of the probabilities
$\xi_E(\lambda)$ weighted according to the composition $\mu_\rho$ of
the state $\rho$.  This is the only agreement demanded here between
quantum mechanics and a hidden-variable theory; that is why we refer
to the resulting no-go theorem as an expectation no-go theorem.

We have deviated here in several ways from Spekkens's formulation in
\cite{spekkens}, and we pause to explain the deviations.  First, while
giving the definition, Spekkens explains ``density operator'' as ``a
positive trace-class operator''.  We take ``density operator'' in its
usual meaning, which requires that the trace of $\rho$ is 1.  We
assume that this is what Spekkens intended, both because of the
terminology and because of the required normalization of $\mu_\rho$.
If $\rho$ were an arbitrary trace-class operator, then we would expect
$\int_\Lambda\mu_\rho(\lambda)\,d\lambda$ to equal the trace of $\rho$
rather than 1.

Second, Spekkens refers to $\Lambda$ as a measurable space rather than
a measure space.  The difference is that a measurable space consists
just of a set $\Lambda$ and a $\sigma$-algebra of subsets called the
measurable sets; a measure space has, in addition, a specific measure
defined on this $\sigma$-algebra.  The integrals in the definition,
both in the normalization condition for $\mu_\rho$ and in the equation
at the end of the definition, presuppose the availability of a fixed
measure denoted by $d\lambda$.  So we assume that Spekkens intended
$\Lambda$ to be a measure space, and we have formulated our definition
accordingly.

Third, we have required the functions $\mu_\rho$ and $\xi_E$ to be
measurable.  This requirement is needed in order for the integrals in
the definition to make sense.

Because the probability densities associated to states (density
operators) $\rho$ are given by functions $\mu_\rho$, they are, when
considered as measures on $\Lambda$, always absolutely continuous with
respect to the fixed measure $d\lambda$.  This aspect of the
definition does not seem well motivated.  It remains in force in
\cite{ferrie2008} and \cite{ferrie2009}, but in the more recent paper
\cite{ferrie2010} it is replaced by a broader viewpoint, taking
$\Lambda$ to be a measurable space (not a measure space, i.e., no
fixed measure) and representing states $\rho$ by measures rather than
by functions.

\begin{rmk} \label{lang} We already mentioned the ontological models
  from \cite{spekkens}; these assign density functions $\mu$ and
  outcome functions $\xi$ to preparations and measurement procedures,
  respectively, rather than to states $\rho$ and effects $E$.  The
  hypotheses for probability representations that we gave above are
  what one obtains by adding to the notion of ontological models the
  additional hypotheses of preparation noncontextuality and
  measurement noncontextuality.  In the same paper \cite{spekkens},
  Spekkens introduces a notion of ``quasiprobability representation'',
  which requires the functions $\mu_\rho$ and $\xi_E$ to be determined
  independently of the preparation of $\rho$ and the apparatus
  measuring $E$, but which allows these functions to take negative
  values.  Thus, our notion of probability representation can be
  obtained by adjoining, to the notion of quasiprobability
  representation, the additional hypothesis of nonnegativity.  In
  other words, the three notions of nonnegative quasiprobability
  representation, noncontextual ontological model (both from
  \cite{spekkens}), and probability representation (in our
  terminology) coincide.  The coincidence of the first two of these
  accounts for the title of \cite{spekkens}.
\end{rmk}

In formulating Definition~\ref{v1}, we have retained one ambiguity
from \cite{spekkens}, namely the third form of noncontextuality,
mentioned in Section~\ref{variants}: Does $\xi_E$ depend only on $E$
or also on the POVM that $E$ is a member of?  The notation $\xi_E$,
which mentions only $E$ and not a whole POVM, suggests the former, but
the wording of the relevant clause in the definition of
``quasiprobability representation'' in \cite{spekkens}, ``every POVM
\dots\ is represented by a set \dots\ of real-valued functions
\dots,'' suggests the latter.  We adopt the former interpretation,
that $\xi_E$ is determined by $E$, for two reasons.  First, the
formulation of measurement noncontextuality in \cite{spekkens}
supports this interpretation.  Second, this interpretation seems to be
essential for the proof of the no-go theorem in \cite{spekkens}.

To complete our discussion of the hypotheses in \cite{spekkens}, one
more assumption needs to be discussed, namely convex-linearity.  This
assumption is not present in the definitions of ``quasiprobability
representation'' and ``ontological model'' nor in the additional
assumptions of nonnegativity and noncontextuality.  It is, however,
explicitly asserted both for density matrices and for effects as if it
were a necessary property of such models.  Specifically, equations (7)
and (8) of \cite{spekkens} say that, for probability distributions
$\{w_j\}$,
\[
\text{if }\rho=\sum_jw_j\rho_j,\text{ then }\mu_\rho(\lambda)=
\sum_jw_j\mu_{\rho_j}(\lambda)
\]
and
\[
\text{if }E=\sum_jw_jE_j,\text{ then }\xi_E(\lambda)=
\sum_jw_j\xi_{E_j}(\lambda).
\]

Spekkens gives a quite plausible argument for the first of these
equations, namely that an ensemble represented by the convex
combination $\rho$ can be prepared by first choosing a value of $j$ at
random, with each $j$ having probability $w_j$, and then preparing the
correponding state $\rho_j$.  The corresponding sub-ensembles should
then be given by the corresponding weighted mixtures of the
sub-ensembles of the $\rho_j$'s.

The plausibility of convex-linearity might be reduced if one considers
the fact that the same $\rho$ can result from such a mixture in many
ways, so convex-linearity imposes some highly nontrivial constraints
on the $\mu$ functions.  Any uneasiness resulting from this
consideration can, however, be ascribed to the assumption of
preparation noncontextuality rather than to convex-linearity.  The
uneasiness results from the requirement that all the many ways to
prepare a $\rho$ ensemble must yield the same mixture of
sub-ensembles.

Despite the plausibility of convex-linearity, it does not follow from
just the definitions in \cite{spekkens} or from our version,
Definition~\ref{v1} above.  To see this, suppose that the functions
$\xi_E$ do not span the whole space of square-integrable functions on
$\Lambda$, so that there is a function $\sigma$ orthogonal to all of
these $\xi_E$'s, where ``orthogonal'' means that
$\int\sigma(\lambda)\xi_E(\lambda)\,d\lambda=0$.  One could then
modify the $\mu_\rho$ functions by adding to each one some multiple of
$\sigma$, obtaining $\mu'_\rho=\mu_\rho+c_\rho\sigma$ and still
satisfying the definitions.  Here the coefficients $c_\rho$ can be
chosen arbitrarily for all of the density operators $\rho$.  By
choosing them in a sufficiently incoherent way, one could arrange that
$\mu'_\rho(\lambda)\neq\sum_jw_j\mu'_{\rho_j}(\lambda)$.

If, on the other hand, the $\xi_E$'s do span the whole space of
functions on $\Lambda$, then Spekkens's desired equation
$\mu_\rho(\lambda)=\sum_jw_j\mu_{\rho_j}(\lambda)$ does follow, for
all but a measure-zero set of $\lambda$'s, because the two sides of
this equation must give the same result when integrated against any
$\xi_E$.

Unfortunately, nothing in the definitions requires the $\xi_E$'s to
span the whole space.  For example, given any probability
representation, we can obtain another, physically equivalent one as
follows.  Replace $\Lambda$ by the disjoint union
$\Lambda_1\sqcup\Lambda_2$ of two copies of $\Lambda$.  Define the
measure of any subset of $\Lambda_1\sqcup\Lambda_2$ to be the average
of the original measures of its intersections with the two copies of
$\Lambda$.  Define all the functions $\mu_\rho$ and $\xi_E$ on the new
space by simply copying the original values on both of the
$\Lambda_i$'s.  The result is a probability representation in
which the $\xi_E$'s span only the space of functions that are the same
on the two copies of $\Lambda$.

Convex-linearity plays an important role in Spekkens's proof of the
no-go theorem in \cite{spekkens}, so, in order to support this proof,
it should be added either as a requirement in the definition of the
probability representations under consideration or as a hypothesis in
the no-go theorem.

Convex-linearity leads to another problem in \cite{spekkens}.
Spekkens asserts that, if a function $f$ is convex-linear on a convex
set \scr S of operators that span the space of Hermitian operators
(and $f$ takes the value zero on the zero operator if the latter is in
\scr S), then $f$ can be uniquely extended to a linear function on
this space.  Unfortunately, such a linear extension need not exist in
the general case, when zero is not in \scr S.\footnote{Spekkens gives
  a formula purporting to define a linear extension of $f$ in general,
  but it is not well-defined because it involves some arbitrary
  choices.  He also gives, in footnote~18 of the newer version
  \cite{spekkens2} of his paper, an argument purporting to show that
  his formula is independent of those choices, but that argument
  fails.  It involves dividing by an appropriate constant $C$ to turn
  two nonnegative linear combinations, the two sides of an equation,
  into convex combinations so that the assumption of convex-linearity
  can be applied.  But the necessary divisor $C$ may need to be
  different for the two sides of the equation.}  For a simple example,
consider the function that is identically 1 on an \scr S that spans
the space of Hermitian operators, does not contain 0, but does contain
two orthogonal projections and their sum.  Because of this difficulty,
we give, in Appendix~\ref{conv-lin}, a careful discussion of
convex-linearity and its relation to linearity.

The no-go theorem in \cite{spekkens} says that, when the Hilbert space
\scr H has dimension at least 2, there cannot be a probability
representation (Spekkens version), subject to the clarifications above, and
satisfying the additional hypothesis of convex-linearity both for
states and for effects.  We give a careful proof of this theorem in
Appendix~\ref{old}.

\subsection{Ferrie and Emerson's no-go theorems}

We turn next to a discussion of the papers
\cite{ferrie2008,ferrie2009} of Ferrie and Emerson.  These papers use
the notion of frames in Hilbert spaces, a generalization of the notion
of basis.  We did not find frames useful, so we describe the relevant
part of these papers in a way that minimizes reference
to frames.

In both \cite{ferrie2008} and \cite{ferrie2009}, a quasiprobability
representation of quantum states\footnote{Not of quantum mechanics but
  merely of quantum states.  A representation of quantum mechanics
  would also include an interpretation for measurements.} is defined
as a linear and invertible map $T$ from the space of Hermitian
operators on \scr H to $L^2(\Lambda,\mu)$.  Here $\Lambda$ is a
measure space\footnote{We use the notation $\Lambda$ for consistency
  with \cite{spekkens} and \cite{ferrie2009}.  The corresponding space
  is called $\Gamma$ in \cite{ferrie2008} and $\Omega$ in
  \cite{ferrie2010}.}, with measure $\mu$, and $L^2(\Lambda,\mu)$ is
the space of real-valued, square-integrable functions on it, modulo
equality $\mu$-almost everywhere.  Note that both $L^2(\Lambda,\mu)$
and the space of Hermitian operators on $\scr H$ are real Hilbert
spaces, the latter having the inner product defined by \tr{AB}.
As far as we can see, the motivation for using $L^2(\Lambda,\mu)$
rather than $L^1(\Lambda,\mu)$ comes neither from intuition nor from
physics but rather from the mathematical benefits of having a Hilbert
space and from the authors' desire to use the frame formalism.

The intuition behind a quasiprobability representation in this sense
is that each $\lambda\in\Lambda$ represents an assignment of possible
values to the hidden variables, or equivalently it represents one of
the sub-ensembles provided by the hidden-variable theory.  For a
density operator $\rho$, the function $T(\rho)$ is the probability
distribution on sub-ensembles in the ensemble described by $\rho$.

We believe that, when requiring $T$ to be invertible, the authors of
\cite{ferrie2008,ferrie2009} meant only to require that it be
one-to-one, not that it be surjective as the usual meaning of
``invertible'' would imply.  In other words, ``invertible'' was
intended to mean merely ``invertible on the range of $T$.''

Comparing the work in these papers with our commentary on
\cite{spekkens} above, we note that in both \cite{ferrie2008} and
\cite{ferrie2009}, the definition of ``density operator'' includes, as
we expected, the requirement that the trace be 1; the space $\Lambda$
is explicitly equipped with a fixed measure $\mu$ (corresponding to
the implicit $d\lambda$ in \cite{spekkens}); and the functions
representing states and effects are required to be measurable.
Because states are represented by functions in the presence of the
fixed measure $\mu$, the probability distributions of the
sub-ensembles within a quantum state's ensemble are always absolutely
continuous with respect to $\mu$, just as in \cite{spekkens}.

Concerning the question whether an effect $E$ completely determines
the function $\xi_E$ or whether $\xi_E$ can depend also on the POVM in
which $E$ occurs, \cite{ferrie2008} contains the same ambiguity as
\cite{spekkens}, but \cite{ferrie2009} unambiguously requires
determinateness here: $\xi_E$ depends only on $E$.

Concerning convex-linearity, the situation in these papers
\cite{ferrie2008,ferrie2009} is rather complicated.  As already
indicated, the definition of a quasiprobability representation of
states in these papers explicitly requires linearity.  For the broader
notion of a quasiprobability representation of quantum mechanics
(incorporating not just states but effects), the discussion in
\cite[Section~3.2]{ferrie2008} begins in the context of frame
representations, which are necessarily linear.  But it continues with
what the authors call a reformulation of the axioms of quantum
mechanics, and this reformulation does not mention convex-linearity.
Indeed, the axioms listed there are very similar to those of Spekkens
\cite{spekkens} that we put into Definition~\ref{v1}.  Just as in our
discussion of \cite{spekkens}, the axioms do not imply
convex-linearity.

In \cite[Section~IV.B]{ferrie2009}, we find a notion of ``frame
representation of quantum theory'' that implies linearity.
Later, in Sections~V.A and V.B, there are notions of ``classical
representation of quantum theory'' and of ``quasi-probability
representation of quantum theory,'' neither of which mentions or
implies convex-linearity.  Lemma~2 in Section~V.B asserts that the
mappings in a quasi-probability representation of quantum theory are
affine, but this lemma is incorrect.  (The error in the proof is the
assumption, in the last displayed implication, that a convex
combination $p\mu_{\sigma_1}+(1-p)\mu_{\sigma_2}$ of two
$\mu$-functions representing states is again such a $\mu$-function, so
that the preceding displayed implication can be applied to it.)

\subsection{Ferrie, Morris, and Emerson's no-go
  theorem}        \label{fme}

The difficulties in \cite{spekkens,ferrie2008,ferrie2009} that we have
pointed out here, are resolved in \cite{ferrie2010}.  In the abstract
and introduction of \cite{ferrie2010}, the authors describe their
contribution as being primarily the extension of the earlier results
in \cite{ferrie2008,ferrie2009} from finite-dimensional Hilbert spaces
to infinite-dimensional ones.  In view of results to be presented in
Section~\ref{reduce} below, showing that in many situations no-go
theorems for one Hilbert space automatically extend to similar
theorems for any larger Hilbert spaces, we regard this extension as
less important than the contribution in \cite{ferrie2010} of giving
precise formulations that correct the deficiencies of prior work.

The properties required of hidden-variable theories in
\cite{ferrie2010} constitute Definition~\ref{v2} below, but before
formulating this definition we need to introduce notations for the
spaces and subsets involved, and we need to point out some
relationships between these spaces.

\begin{notat}   \phantomsection        \label{what}
\begin{ls}
\item  In the following, let $\Lambda$ be a measurable space.  Recall
    that this means that the set $\Lambda$ is equipped with a
    specified $\sigma$-algebra $\Sigma$ of subsets.
\item $\scr F(\Lambda,\Sigma)$, often abbreviated to simply $\scr F$,
  is the space of bounded, measurable, real-valued functions on
  $\Lambda$.  It is a vector space over the real numbers, and we equip
  it with the supremum norm, $\nm
  f=\sup\{|f(\lambda)|:\lambda\in\Lambda\}$.
\item $\scr F_{[0,1]}(\Lambda,\Sigma)$ or simply $\scr F_{[0,1]}$ is
  the subset of \scr F consisting of those functions whose values lie
  in the interval $[0,1]$.
\item $\scr M(\Lambda,\Sigma)$, often abbreviated to simply \scr M, is
  the space of bounded, signed, real-valued measures on $\Lambda$.  It
  is a vector space over the real numbers, and we equip it with the
  total variation norm.  That is, if $\mu\in\scr M$, then $\mu$ can be
  expressed as $\mu_+-\mu_-$, where $\mu_+$ and $\mu_-$ are positive
  measures with disjoint supports (called the positive and negative
  parts of $\mu$). Then
  $\nm\mu=\mu_+(\Lambda)+\mu_-(\Lambda)$.
\item $\scr M_{+1}(\Lambda,\Sigma)$ or simply $\scr M_{+1}$ is the
  subset of \scr M consisting of the probability measures, i.e., the
  positive measures with total measure equal to 1.
\item \scr H is a complex Hilbert space.
\item $\scr B(\scr H)$, often abbreviated to simply \scr B, is the
  real Banach space of bounded, self-adjoint operators $\scr H\to\scr
  H$; its norm is the operator norm $\nm A=\sup\{\nm{Ax}:x\in
  A,\,\nm x=1\}$.
\item $\scr B_{[0,1]}(\scr H)$ or simply $\scr B_{[0,1]}$ is the
  subset of \scr B consisting of the effects, i.e., operators
  $A\in\scr B$ such that both $A$ and $I-A$ are positive\footnote{$I$
    is the identity operator. Positivity of a self-adjoint operator
    $A$ means that all of its spectrum lies in the non-negative half
    of the real line.  Equivalently, it means that $\bra\psi
    A\ket\psi\geq0$ for all $\ket\psi\in\scr H$}, or equivalently such
  that the spectrum of $A$ lies in the interval $[0,1]$.
\item $\scr T(\scr H)$, often abbreviated simply \scr T, is the vector
  subspace of \scr B consisting of the (self-adjoint) trace-class
  operators.  These are the operators $A$ whose spectrum consists only
  of (real) eigenvalues $\alpha_i$ (eigenvalues with multiplicity $>1$
  are repeated in this list; the continuous spectrum is empty or
  $\{0\}$) such that the sum $\sum_i|\alpha_i|$ is finite; this sum
  serves as the norm \nm A of $A$ in \scr T.  (Note that this norm is
  usually not equal to the operator norm, the norm of $A$ in \scr B,
  which equals the supremum of the $|\alpha_i|$.)
\item $\scr T_{+1}(\scr H)$ or simply $\scr T_{+1}$ is the subset of
  \scr T consisting of the density operators, positive operators of
  trace 1.
  \end{ls}
\end{notat}

\begin{rmk}
 We have modified some of the notations from \cite{ferrie2010}. In
  the first place, we have removed a subscript $s$ from \scr T and
  \scr B.  The subscript's purpose was to indicate that these spaces
  consist only of self-adjoint operators. Since we do not deal with
  more general operators in this context, the subscript seemed
  superfluous.  Also, what we have called $\scr F_{[0,1]}$, $\scr
  M_{+1}$, $\scr B_{[0,1]}$, and $\scr T_{+1}$ have in
  \cite{ferrie2010} the notations $\scr E(\Lambda,\Sigma)$, $\scr
  S(\Lambda,\Sigma)$, $\scr E(\scr H)$, and $\scr S(\scr H)$,
  respectively.  The double use of \scr E and \scr S served the useful
  purpose of indicating which ingredients of quantum theory correspond
  to which ingredients of a hidden-variable theory, but they also
  prevented any abbreviations omitting $(\Lambda,\Sigma)$ or \scr H.
  We hope that our notations will be easier to remember, since the
  main symbols (\scr{F,M,B,T}) indicate the vector spaces in which
  these subsets lie, while the subscripts hint at the restriction that
  characterizes elements of the subset.
\end{rmk}

In contrast to \cite{spekkens,ferrie2008,ferrie2009} there is no
specified measure on $\Lambda$.  As in these papers discussed earlier,
a point $\lambda\in\Lambda$ represents specific values for all the
hidden variables, and thus represents a specific sub-ensemble for the
hidden-variable theory.  A quantum state will then be viewed as a
mixture of such sub-ensembles according to a probability measure on
$\Lambda$, i.e., an element of $\scr M_{+1}$.  This approach avoids
any assumption of absolute continuity of these measures with respect
to an a priori given measure; there simply is no a priori given
measure.

The normed vector spaces introduced above are connected by two duality
relations.  First, every $f\in\scr F$ induces a continuous
linear functional $\bar f$ on  \scr M by integration:
\[
\bar f(\mu)=\int_\Lambda f\,d\mu.
\]
We shall not need to deal with the entire dual space%
\footnote{In general, and even for nice measurable spaces like the
  real line \bbb R with the $\sigma$-algebra of Borel sets, $\scr M'$
  is an unpleasantly complicated space.  In particular, in this
  special case of \bbb R, the linear functional assigning to each
  measure $\mu\in\scr M$ the total measure of all the individual
  points, $\sum_{x\in\mathbb R}\mu(\{x\})$, is not of the form $\bar
  f$ for any $f\in\scr F$.  For more information about the dual of
  \scr M, see, for example, \cite{mathse} and the references cited
  there.}%
$\scr M'$ of \scr M but only with the part consisting of functionals
$\bar f$ arising from \scr F.

Second, the dual $\scr T'$ of \scr T can be identified with \scr B as
follows.  Every $B\in\scr B$ induces a continuous linear functional
$\bar B$ on \scr T by
\[
\bar B(W)=\tr{BW},
\]
because the product of a bounded operator and a trace-class operator
is again in the trace class (i.e., the trace class is an ideal in the
ring of bounded operators).  Furthermore, every bounded linear
functional on \scr T arises in this way from a unique $B\in\scr B$.
The correspondence $B\mapsto\bar B$ is an isometric isomorphism
between \scr B and $\scr T'$, and so one often identifies these two
spaces.  For details about this duality, see, for example,
\cite[Theorem~23]{skoufranis}.

\begin{rmk}             \label{asymmety}
Note that this duality relationship is not symmetric.  That is,
although each $W\in\scr T$ induces a continuous linear functional on
\scr B, namely $A\mapsto\tr{AW}$, these will not be all of the linear
functionals on \scr B unless \scr H is finite-dimensional.

Spekkens \cite{spekkens} emphasizes a certain symmetry between states
and measurements, and, at the end of the paper, he seeks to give an
``even-handed'' proof of a no-go theorem, respecting this symmetry.
The fact that \scr B, the space in which measurements live, is the
dual of \scr T, the space in which states live, but not vice versa,
suggests that the actual situation is not really symmetrical.

One reflection of this asymmetry arises when we try to prove a no-go
theorem for probability representations (Spekkens version) as defined
above.  After building into that definition our clarifications and
corrections of Spekkens's assumptions, the proof that we obtained, and
which we record in Appendix~\ref{old} below, is not even-handed in the
sense desired by Spekkens.  We do not have any even-handed proof of an
expectation no-go theorem.

The asymmetry in the duality relationship between \scr B and \scr T is
specific to the case of infinite-dimensional spaces.  In the case of
finite-dimensional \scr H, say of dimension $d$, all bounded linear
operators are in the trace class, so $\scr B$ and \scr T are the same
when considered just as vector spaces.  Their norms, though not
identical, are equivalent in the sense that each is bounded by a
constant (depending on $d$) multiple of the other.  They are
identified with the space of Hermitian $d\times d$ matrices.

To get a really smooth symmetry, though, one would need not only that
\scr H is finite-dimensional but also that $\Lambda$ is finite.  That
additional finiteness would make \scr M and \scr F dual to each other
and would avoid the messiness that arises in $\scr M'$ in the general
case.  Unfortunately, finiteness of $\Lambda$ is quite a restrictive
assumption.  Consider, for example, a spin-$\frac12$ particle in an
eigenstate of the $z$-spin.  The hidden variables in this situation
would have to determine the spin components in all directions other
than $z$, and there is a continuum of possibilities there.  It seems
that finiteness of $\Lambda$ becomes plausible only if one can argue
that, because of limited precision of measurements, the spaces of
measurement outcomes can be discretized and thus treated as finite.

See Section~\ref{example} below for further discussion of symmetry (or
its absence) in the light of some specific examples.

\end{rmk}

We are now in a position to present the notion that Ferrie et
al. \cite{ferrie2010} call a \emph{classical representation of quantum
  mechanics}.  We prefer to call it a probability representation,
viewing it as an updating and clarification of the notion introduced
in Definition~\ref{v1}.

\begin{df}               \label{v2}
A \emph{probability representation (Ferrie-Morris-Emerson version)}
  for quantum systems described by \scr H consists of
  \begin{ls}
\item a measurable space $\Lambda$,
\item a convex-linear map $T$ from the set $\scr T_{+1}$ of density
  matrices into the set $\scr M_{+1}$ of probability measures, and
\item a convex-linear map $S$ from the set $\scr B_{[0,1]}$ of effects
  into the set $\scr F_{[0,1]}$ of measurable functions from $\Lambda$
  to $[0,1]$,
\end{ls}
subject to, for all $\rho\in\scr T_{+1}$ and all $E\in\scr B_{[0,1]}$,
\[
\tr{\rho E}=\int_\Lambda S(E)\,dT(\rho).
\]
\end{df}

The correspondence between this definition and the earlier
Definition~\ref{v1} is that the measure $T(\rho)$ is what was
previously written $\mu_\rho(\lambda)\,d\lambda$, and $S(E)$ was
previously $\xi_E$.  The ``trace equals integral'' requirement in the
last clause of the definition still says that the expectation of the
effect $E$ in the state $\rho$ is the same whether computed in quantum
mechanics (the trace) or in the hidden-variable theory (the integral).

Theorem~1 of \cite{ferrie2010} asserts that such a probability
representation is impossible (provided \scr H has dimension at least
2).  The proof has a gap, which we fill in the next section, and we
simultaneously make some other improvements to the theorem and its
proof.

\subsection{Our Expectation No-Go Theorem}

In this section, we prove the first main result of this paper, a no-go
theorem that strengthens Theorem~1 of \cite{ferrie2010}. Our theorem
and its proof are based on the result in \cite{ferrie2010} but differ
from it in two major respects.  First, we use a weaker hypothesis,
requiring the existence of $S(E)$ only for (certain) sharp effects
$E$, not for all effects.  We impose no convex-linearity assumption on
$S$. Second, we fill a gap that apparently resulted from quoting a
misstated fact in \cite{bugajski}.  In addition to these changes, we
also remove an unnecessary paragraph in the otherwise terse proof.

The following definition, our final updating of the notion of
``probability representation,'' expresses the hypotheses necessary for
our theorem.  The conventions in Notation~\ref{what} remain in
force.

\begin{df}               \label{v3}
A \emph{probability representation (our version)}
  for quantum systems described by \scr H consists of
  \begin{ls}
\item a measurable space $\Lambda$,
\item a convex-linear map $T$ from the set $\scr T_{+1}$ of density
  matrices into the set $\scr M_{+1}$ of probability measures, and
\item a map $S$ from the set of rank-1 projections in \scr H
  into the set $\scr F_{[0,1]}$ of measurable functions from $\Lambda$
  to $[0,1]$,
  \end{ls}
  subject to, for all $\rho\in\scr T_{+1}$ and all rank-1 projections
  $E$,
\[
\tr{\rho E}=\int_\Lambda S(E)\,dT(\rho).
\]

\end{df}

This definition differs from the previous version,
Definition~\ref{v2}, in that the domain of $S$ is no longer the set
$\scr B_{[0,1]}$ of all effects but the much smaller set of sharp
effects of rank 1.  The requirement that $S$ be convex-linear is
removed, because it would make no sense when the domain of $S$ is not
convex.

\begin{rmk}
  The restriction to sharp effects is significant because, as
  explained in Remark~\ref{eff-op}, measuring an effect $E$ is not in
  general the same as measuring the observable that is given by the
  same self-adjoint operator $E$.  The two sorts of measurement are
  the same if and only if $E$ is a sharp effect, i.e., a projection
  operator from \scr H to a closed subspace.  Thus, sharp effects are
  the area common to the effect-based hidden-variable notions
  considered in this section and the observable-based hidden-variable
  theories to be discussed in Section~\ref{val} below.

Definition~\ref{v3} reduces the domain of $S$ not only to the set of
sharp effects but to the even smaller set of projections for which the
rank, the dimension of the range, is 1.  This additional reduction is
included simply as a mathematical optimization of the theorem.
\end{rmk}

Since the domain of $S$ is, in Definition~\ref{v3}, no longer a convex
set, there is no requirement that $S$ be convex-linear.  In principle,
a quite arbitrary function could serve as $S$, though, as we shall see
in the proof of the theorem below, the last clause of the definition,
equating a trace to an integral, implies a remnant of linearity for
$S$, namely that $S$ is one-to-one and its inverse is the restriction
to the range of $S$ of a linear transformation.

We now turn to our expectation no-go theorem,
Theorem~\ref{exp-thm-intro} in the introduction, expressing it in the
language of probability representations.

\begin{thm}     \label{exp-thm}
  If the Hilbert space \scr H has dimension at least 2, then there is
  no probability representation (our version) for quantum systems
  described by \scr H.
\end{thm}

\begin{pf}
Suppose, toward a contradiction, that we have a probability
representation (our version), consisting of $\Lambda,T,S$, for some \scr
H of dimension at least 2.
  We begin by working with the convex-linear map $T:\scr T_{+1}\to\scr
  M_{+1}$, and our first objective is to extend it to a linear map,
  still called $T$, from all of \scr T into \scr M.  For general
  information about such extensions of convex-linear maps, see
  Appendix~\ref{conv-lin}, but for the case at hand it is convenient
  to give the following very specific argument.  Any trace-class
  self-adjoint operator $A\in\scr T$ can be written as the difference
  of two positive trace-class operators $A=A_+-A_-$, where $A_+$ has
  the same positive eigenvalues and corresponding eigenspaces as $A$
  but is identically zero on all the eigenspaces corresponding to
  non-positive eigenvalues.  $-A_-$ similarly matches the negative
  eigenvalues and eigenspaces of $A$; we reverse its sign to get the
  positive operator $A_-$.  As long as neither $A_+$ nor $A_-$ is
  zero, we can multiply them by suitable scalars to produce operators
  with trace 1, i.e., elements of $\scr T_{+1}$, and thus we can write
  $A=bB-cC$ where $B,C\in\scr T_{+1}$ and $b,c$ are positive real
  numbers.  If one or both of $A_+$ and $A_-$ is zero, then we still
  have such a formula for $A$ but one or both of $b$ and $c$ will be
  zero.  So we always have $A=bB-cC$ where $B,C\in\scr T_{+1}$ and
  $b,c\geq0$.  Note for future reference that in this situation $\tr
  A=b-c$ and, for the particular construction of $A_\pm$, $B$, and $C$
  given here, $\nm A=b+c$.  (The norm here is that in \scr T, which we
  defined as the sum of the absolute values of the eigenvalues.)

We extend $T$ to a map $T:\scr T\to\scr M$ by setting, with notation
as above, $T(A)=bT(B)-cT(C)$.  Even though $A$ can have many
representations as $bB-cC$ with $B,C\in\scr T_{+1}$ and $b,c\geq0$,
they all yield the same $T(A)$. Indeed, if $b'B'-c'C'$ is another such
representation, then from $bB-cC=A=b'B'-c'C'$, we obtain
$bB+c'C'=b'B'+cC$. Furthemore, since all of $B,C,B',C'$ have trace
1, we also have $b+c'=b'+c$, and therefore
\[
\frac b{b+c'}B+\frac{c'}{b+c'}C'=\frac{b'}{b'+c}B'+\frac c{b'+c}C.
\]
Here, both sides are convex combinations, so convex-linearity of $T$
yields
\[
\frac b{b+c'}T(B)+\frac{c'}{b+c'}T(C')=
\frac{b'}{b'+c}T(B')+\frac c{b'+c}T(C).
\]
Transposing some terms and clearing fractions (remembering that
$b+c'=b'+c$), we get
\[
bT(B)-cT(C)=b'T(B')-c'T(C'),
\]
which means that $T(A)$ is well-defined.  An easy computation then
shows that $T$ is linear.

We claim that $T$ is a bounded linear transformation.  To this end,
consider some $A$ with $\nm A\leq1$ in \scr T.  Then, as indicated
above, we can represent $A$ as $bB-cC$ with $B,C\in\scr T_{+1}$, with
$b,c\geq0$, and with $b+c=\nm A\leq 1$.  Now $T(B)$ and $T(C)$ are
measures with norm 1 in \scr M.  So $T(A)=bT(B)-cT(C)$ has norm at
most $b+c\leq 1$.  This completes the proof that $T:\scr T\to\scr M$
is a bounded linear transformation.

It follows that $T$ induces a bounded linear transformation on the
dual spaces, $T':\scr M'\to\scr T'$.  In detail, $T'$ sends any
bounded linear functional $h\in\scr M'$ (which means $h:\scr M\to\bbb
R$) to the bounded linear functional $T'(h)=h\circ T:\scr T\to\bbb R$;
\[
T'(h)(A)=h(T(A))\quad\text{for all }h\in\scr M'\text{ and all
}A\in\scr T.
\]

Recall from the discussion in Subsection~\ref{fme} how the dual space
$\scr T'$ of \scr T is identified with \scr B and part of the dual
space $\scr M'$ of \scr M is identified with \scr F.  Via these
identifications, $T':\scr M'\to\scr T'$ restricts to a bounded linear
transformation, which we still call $T'$, from \scr F to \scr B.
Untangling the definitions, we find that, for each $f\in\scr F$,
$T'(f)$ is the unique element of \scr B that satisfies
\begin{equation}
  \label{eq:1}
\tr{T'(f)A}=\int_\Lambda f\,dT(A)\quad \text{for all }A\in\scr T.
\end{equation}
Indeed, the left side of this equation is the value obtained by
applying to $A\in\scr T$ the functional identified with $T'(f)\in\scr
B$, while the right side is the value obtained by applying to the
measure $T(A)$ the functional identified with $f\in\scr F$.

Note also that this equation, though true for all $A\in\scr T$, would
still suffice to uniquely determine $T'(f)$ if it were asserted only
for $A\in\scr T_{+1}$; this is because, as we showed above, the linear
span of $\scr T_{+1}$ is the whole space \scr T.

We now invoke the last clause in Definition~\ref{v3} to find that, for
all rank-1 projections $E$ and all $\rho\in\scr T_{+1}$,
\[
\tr{E\rho}=\int_\Lambda S(E)\,dT(\rho)=\tr{T'(S(E))\rho)}.
\]
But this is, as we saw in the preceding paragraph, enough to show that
$T'(S(E))=E$.

Recall that we imposed no linearity conditions on $S$. Nevertheless,
because $T'$ is linear, this last equation gives what can be viewed as
a weak linearity requirement for $S$.  On its range, $S$ is inverted
by a linear transformation $T'$.

So far, we have followed the argument in \cite{ferrie2010} fairly
closely, just adding some details, for example the reason why $T$ is
bounded, and noting that a drastically reduced domain of $S$ suffices.
At this point, though, Ferrie et al.\ claim, quoting Bugajski
\cite{bugajski}, that the linearity of $T'$ implies that it preserves
a property called coexistence.  Unfortunately, this preservation claim
needs not only that $T'$ is linear but also that it preserves
positivity and sends the constant function 1 to the identity operator.
$T'$ actually has these properties, but this needs to be checked; we
give the proof below.  Also, although we could work with the general
notion of coexistence, it turns out to be more convenient to use an
equivalent formulation, from \cite{teiko1}, for the special case of
two effects.  (For readers interested in the general notion, we
suggest \cite{bugajski} and \cite{teiko1}.)

In preparation for the next step in the proof, we need some
computations.  The first of these is to compute $T'(1)$, where
$1\in\scr F$ means the constant function with value 1.  Referring to
the formula \eqref{eq:1} characterizing $T'$ and remembering that it
suffices to have this formula for $A\in\scr T_{+1}$, we see that
$T'(1)$ is the unique bounded linear operator that satisfies, for all
$\rho\in\scr T_{+1}$,
\[
\tr{T'(1)\rho}=\int_\Lambda\,dT(\rho)=T(\rho)(\Lambda)=1=\tr\rho
=\tr{I\rho},
\]
where the third equality comes from the fact that $T$ maps $\scr
T_{+1}$ into the space $\scr M_{+1}$ of probability measures. Thus,
$T'(1)=I$.

The other computation that we need is conveniently summarized in the
following lemma. Recall that a bounded linear operator $A$ is said to be
positive if $\bra\psi A\ket\psi\geq0$ for all $\ket\psi\in\scr H$ and
that $A\leq B$ means that $B-A$ is positive.

\begin{la} \label{la:pos} If $f\in\scr F$ is nonnegative (meaning
  $f(\lambda)\geq0$ for all $\lambda\in\Lambda$), then $T'(f)$ is a
  positive operator.  Therefore, if $f\leq g$ pointwise in \scr F then
  $T'(f)\leq T'(g)$ in \scr B.
\end{la}

\begin{pf}
  The second assertion follows immediately from the first applied to
  $g-f$, because $T'$ is linear.  To prove the first assertion,
  suppose $f\in\scr F$ is nonnegative, and let $\ket\psi$ be any
  vector in \scr H. The conclusion we want to deduce, $\bra\psi
  T'(f)\ket\psi\geq0$, is obvious if $\ket\psi=0$, so we may assume
  that \ket\psi\ is a non-zero vector.  Normalizing it, we may assume
  further that its length is 1.  Then $\ket\psi\bra\psi\in\scr T_{+1}$
  and therefore $T(\ket\psi\bra\psi)\in\scr M_{+1}$. Using
  equation~\eqref{eq:1}, we compute
\[
\bra\psi T'(f)\ket\psi=\tr{T'(f)\ket\psi\bra\psi}
=\int_\Lambda f\,dT(\ket\psi\bra\psi)\geq0,
\]
where we have used that both the measure $T(\ket\psi\bra\psi)$ and
the integrand $f$ are nonnegative.\footnote{The proof would break down
  here if we were working with possibly negative quasiprobabilities.}
\end{pf}

The following lemma says, in view of a criterion of Heinosaari
\cite[equation~(2)]{teiko1}, that any two elements of $\scr F_{[0,1]}$
coexist.

\begin{la}      \label{la:coexist}
  If $f,g\in\scr F_{[0,1]}$, then there exists $h\in\scr F_{[0,1]}$
  such that all four of $h$, $f-h$, $g-h$, and $1-f-g+h$ are
  nonnegative.
\end{la}

\begin{pf}
  Define $h(\lambda)=\min\{f(\lambda),g(\lambda)\}$ for all
  $\lambda\in\Lambda$.  Then the first three of the assertions in the
  lemma are obvious, and the fourth becomes obvious if we observe that
  $f+g-h=\max\{f,g\}\leq 1$.
\end{pf}

\begin{cor}     \label{cor:coexist}
  For any two rank-1 projections $A,B$ of \scr H, there exists an
  operator $H\in\scr B$ such that all four of $H$, $A-H$, $B-H$, and
  $I-A-B+H$ are positive operators.
\end{cor}

\begin{pf}
  Apply Lemma~\ref{la:coexist} with $f=S(A)$ and $g=S(B)$, let $h$ be
  the function given by the lemma, and let $H=T'(h)$.  The
  nonnegativity of $h$, $f-h$, $g-h$, and $1-f-g+h$ implies, by
  Lemma~\ref{la:pos}, the positivity of $T'(h)=H$, $T'(S(A)-h)=A-H$,
  $T'(S(B)-h)=B-H$, and $T'(1-S(A)-S(B)+h)=I-A-B+H$, where we have
  also used the linearity of $T'$, the fact that $T'(1)=I$, and the
  formula $T'(S(A))=A$ for all $A$ in the domain of $S$.
\end{pf}

Let us apply this corollary to two specific rank-1 projections.  Fix
two orthonormal vectors \ket0 and \ket1. (This is where we use that
\scr H has dimension at least 2.)  Let $\ket+=(\ket0+\ket1)/\sqrt2$.
We use the projections $A=\ket0\bra0$ and $B=\ket+\bra+$ to the
subspaces spanned by $\ket0$ and $\ket+$.  Let $H$ be as in
Corollary~\ref{cor:coexist} for these projections $A$ and $B$.

From the positivity of $H$ and of $A-H$, we get that
$0\leq\bra1H\ket1$ and that
\[
0\leq\bra1(A-H)\ket1=\bra1A\ket1-\bra1H\ket1=-\bra1H\ket1,
\]
where we have used that \ket1, being orthogonal to \ket0, is
annihilated by $A$.  Combining the two inequalities, we infer that
$\bra1H\ket1=0$ and therefore, since $H$ is positive, $H\ket1=0$.
Similarly, using the orthogonal vectors $\ket+$ and
$\ket-=\ket0-\ket1)/\sqrt2$ in place of \ket0 and \ket1, we obtain
$H\ket-=0$.  So, being linear, $H$ is identically zero on the subspace
of \scr H spanned by \ket1 and \ket-; note that \ket0 is in this
subspace, so we have $H\ket0=0$.

Now we use the part of Corollary~\ref{cor:coexist} that has not yet
been used, namely the positivity of $I-A-B+H$.  Since $H\ket0=0$, we
can compute
\[
0\leq\bra0(I-A-B+H)\ket0=\sq{0|0}-\bra0A\ket0-\bra0B\ket0=
1-1-\frac1{\sqrt2}=\frac{-1}{\sqrt2}.
\]
This contradiction completes the proof of the theorem.
\end{pf}

\section{Value No-Go Theorems} \label{val}

We turn now to a different species of no-go theorems, ones saying that
hidden-variable theories cannot even produce the correct outcomes for
individual measurements, let alone the correct probabilities or
expectation values.  Such theorems considerably predated the
expectation no-go theorems considered in the preceding section.  Value
no-go theorems were first established by Bell \cite{bell64,bell66} and
then by Kochen and Specker \cite{ks}; we shall also refer to the
user-friendly exposition given by Mermin \cite{mermin}.

Note that there is no implication in either direction between value
no-go theorems and expectation no-go theorems.  The former say that a
hidden-variable theory cannot predict the correct values for measured
quantities, but it might still predict the correct expectations; the
latter say that a hidden-variable theory cannot predict the correct
expectations, but it might still predict the correct values.

Of course, in order to formulate value no-go theorems, one must
specify what ``correct outcomes for individual measurements'' means.
For this purpose, we need the notion of the joint spectrum of
commuting operators on Hilbert space, and we devote the next
subsection to summarizing the basic facts about joint spectra.

\subsection{Joint Spectra}

A general reference for the notion of joint spectrum is
\cite[Section~6.5]{spectral}.

Let $A_1,\dots,A_n$ be a finite list of pairwise commuting,
self-adjoint operators on a Hilbert space \scr H.  The notion of the
joint spectrum of such a list is a natural generalization of the
notion of the spectrum of a single self-adjoint operator.

The simplest case occurs when the operators are simultaneously
diagonalizable, i.e., when \scr H admits an orthonormal basis
consisting of common eigenvectors of all the $A_i$'s.  In this case,
the joint spectrum consists of the $n$-tuples of scalars
$\nu=(\nu_1,\dots,\nu_n)\in\bbb R^n$ that occur as the eigenvalues
for such common eigenvectors.  That is, $\nu$ belongs to the joint
spectrum if and only if there is a non-zero vector $\ket\psi\in\scr H$
such that $A_i\ket\psi=\nu_i\ket\psi$ for $i=1,\dots,n$.

If \scr H is finite-dimensional, then this simple case is the only one
that can arise, but for infinite-dimensional \scr H we must take into
account the possibility of a continuous spectrum (instead of, or in
addition to, the discrete spectrum given by eigenvectors).  A point
$\nu\in\bbb R^n$ belongs to the joint spectrum $\sigma(A_1,\dots,A_n)$
of $A_1,\dots,A_n$ if and only if it is approximately a tuple of
eigenvalues in the following sense: For every positive $\eps$, there
is a unit vector $\ket\psi\in\scr H$ (an approximate simultaneous
eigenvector) such that, for each $i=1,\dots,n$, we have
$\nm{A_i\ket\psi-\nu_i\ket\psi}<\eps$.

The joint spectrum of a tuple of self-adjoint operators is a closed
subset of $\bbb R^n$.  If the operators are bounded, then so is their
joint spectrum.

Just as for a single operator, there is a spectral decomposition
leading to a functional calculus for tuples of commuting self-adjoint
operators.  In more detail, there is a unique spectral measure $E$, a
countably additive map from Borel subsets of $\bbb R^n$ to projection
operators on \scr H, such that, for each $i$,
\[
A_i=\int_{\bbb R^n} x_i\,dE(x_1,\dots,x_n).
\]

The joint spectrum $\sigma(A_1,\dots,A_n)$ can be characterized as the
support of this spectral measure, i.e., the set of points $\nu\in\bbb
R^n$ such that $E(B)\neq0$ for all neighborhoods $B$ of $\nu$.

The preceding information about joint spectra is explicit in
\cite[Section~6.5]{spectral}.  (For the boundedness of the joint
spectrum of commuting bounded operators, look at the proof of
Theorem~1 in that section.)  What follows is implicit in the
statement, on page~155 of \cite{spectral}, that most of Section~1,
Subsection~4, which concerns functions of a single operator, can be
repeated in the present context of several commuting operators.  We
fill in some arguments that are not given in that subsection of
\cite{spectral}.

Given a Borel function $f:\bbb R^n\to\bbb R$, one defines
\[
f(A_1,\dots,A_n)=\int_{\bbb R^n} f(x_1,\dots,x_n)\,dE(x_1,\dots,x_n).
\]
We shall use this notion only for continuous $f$, and in this case we
have the following useful information.

\begin{prop}            \label{jspec}
  Let $A_1,\dots,A_n$ be commuting, self-adjoint operators, with joint
  spectrum $\sigma(A_1,\dots,A_n)$.  Then, for any continuous $f:\bbb
  R^n\to\bbb R$, we have $f(A_1,\dots,A_n)=0$ if and only if $f$
  vanishes identically on $\sigma(A_1,\dots,A_n)$.  Furthermore, a
  point $\nu\in\bbb R^n$ belongs to $\sigma(A_1,\dots,A_n)$ if and
  only if every continuous function $f:\bbb R^n\to\bbb R$ that
  satisfies   $f(A_1,\dots,A_n)=0$  also satisfies $f(\nu)=0$.
\end{prop}

\begin{pf}
Although we have two ``if and only if'' statements to prove, their
``only if'' halves say the same thing, so we need only to prove three
implications:
\begin{lsnum}
\item If a continuous function $f:\bbb R^n\to\bbb R$ vanishes
  identically on $\sigma(A_1,\dots,A_n)$, then $f(A_1,\dots,A_n)=0$.
\item If $f(A_1,\dots,A_n)=0$ for a continuous $f$ and if
  $\nu\in\sigma(A_1,\dots,A_n)$, then $f(\nu)=0$.
\item If $\nu\notin\sigma(A_1,\dots,A_n)$, then there is a continuous
  $f:\bbb R^n\to\bbb R$ with $f(A_1,\dots,A_n)=0$ but $f(\nu)\neq 0$
\end{lsnum}

Item~(1) here is clear from the definition of $f(A_1,\dots,A_n)$.  It
is the integral of $f$ with respect to $E$, and $f$ vanishes on the
support of $E$.

For item~(2), we use the generalization to several commuting operators
of a fact from the cited subsection of \cite{spectral}, namely that
\[
  \nm{f(A_1,\dots,A_n)}=E\text{-}\sup\{|f(\nu)|:\nu\in
  \sigma(A_1,\dots,A_n)\}.
\]
Here the notation $E\text{-}\sup$ means the essential supremum with
respect to the spectral measure $E$, which is the infimum of all the
numbers $a$ such that $E(\{\nu:|f(\nu)|>a\})=0$.  In the situation of
item~(2), we therefore have that this essential supremum is zero.
Suppose now, toward a contradiction, that
$\nu\in\sigma(A_1,\dots,A_n)$ is a point for which $f(\nu)\neq0$.
Since $f$ is continuous and $f(\nu)\neq0$, there is an open
neighborhood $N$ of $\nu$ such that, for all $x\in N$,
$|f(x)|>\frac12|f(\nu)|>0$.  Since the essential supremum of $|f|$ is
zero, there is an $a<\frac12|f(\nu)|$ for which $E(\{x:|f(x)|>a\})=0$.
But the set $\{x:|f(x)|>a\}$ includes $N$, so $E(N)=0$.  This is a
contradiction, because every neighborhood $N$ of a point $\nu$ in the
joint spectrum must have $E(N)\neq0$.

Finally, to prove item~(3), suppose $\nu\notin\sigma(A_1,\dots,A_n)$
and notice that, thanks to item~(1), we need only find a continuous
$f$ that vanishes identically on $\sigma(A_1,\dots,A_n)$ but does not
vanish at $\nu$.  Since $\sigma(A_1,\dots,A_n)$ is closed, the
function sending each point in $\bbb R^n$ to its distance from
$\sigma(A_1,\dots,A_n)$ does the job.
\end{pf}

The last assertion in Proposition~\ref{jspec} can be summarized as:
The joint spectrum of $A_1,\dots,A_n$ consists of all
those points $(\nu_1,\dots,\nu_n)$ that satisfy all the same equations
as the operators themselves.  Here ``equations'' should be understood
as equations between continuous functions.

Just as the points in the spectrum of a single Hermitian operator $A$
are, according to quantum theory, the possible results of a
measurement of $A$, so the points in the joint spectrum of
$A_1,\dots,A_n$ are the possible outcomes of a simultaneous
measurement of all of $A_1,\dots,A_n$.  Note that both mathematics and
physics require the operators $A_1,\dots,A_n$ here to commute ---
mathematics in order that the joint spectrum be defined, and physics
in order that these observables be simultaneously measurable.

We record, for future reference, some very special cases of the
definition of joint spectrum.  These all fall under the simple case
mentioned at the beginning of this subsection: the operators will be
simultaneously diagonalizable, so the joint spectrum consists of the
eigenvalues for the common eigenvectors of the operators
$A_1,\dots,A_n$.  If the $A_i$ are projections, then each point in
their joint spectrum is a tuple of zeros and ones.  If $A_1,\dots,A_n$
are the rank-1 projections to an orthogonal set of directions, then
their joint spectrum contains all the $n$-tuples consisting of a
single one and $n-1$ zeros.  The only other point that could be in
the joint spectrum is the $n$-tuple of all zeros; it is present if and
only if the directions to which that $A_i$'s project do not span the
whole space \scr H.

\subsection{Value Maps}

Now we are ready to define precisely what is expected of a
hidden-variable theory in order for it to predict the correct values
for observables. The following definition, which is based on the
discussion in \cite[Section~II]{mermin}, is intended to provide that
specification.

\begin{df} \label{valmap} Let \scr H be a Hilbert space, and let \scr
  O be a set of observables, i.e., self-adjoint operators on \scr H.
  A \emph{value map} for \scr O in \scr H is a function $v$ assigning
  to each observable $A\in\scr O$ a number $v(A)$ in the spectrum of
  $A$, in such a way that, whenever $A_1,\dots,A_n$ are pairwise
  commuting elements of \scr O, then $(v(A_1),\dots,v(A_n))$ is in the
  joint spectrum of $(A_1,\dots,A_n)$.
\end{df}

The intention behind this definition is that, in a hidden-variable
theory, a quantum state represents an ensemble of individual systems,
each of which has definite values for observables. That is, each
individual system has a value map associated to it, describing what
values would be obtained if we were to measure observable properties
of the system.  A believer in such a hidden-variable theory would
expect a value map for the largest possible \scr O, the set of all
self-adjoint operators on \scr H, unless there were superselection
rules rendering some such operators unobservable.

The part of Definition~\ref{valmap} about pairwise commuting operators
says exactly that, if one measures the observables $A_1,\dots,A_n$
simultaneously, which is possible because they commute, then the values
one obtains should be among the possibilities permitted by quantum
mechanics, namely the $n$-tuples in the joint spectrum of the
operators.

On the other hand, for observables that do not
commute, quantum mechanics does not allow them to be simultaneously
exactly measured, does not describe possible simultaneous values, and
thus does not impose restrictions on value maps.

\subsection{No-Go Theorem}

A hidden-variable theory should do more than just provide some value
maps describing the properties of the sub-ensembles inside the quantum
states. It should provide, for each quantum state $\rho$, a
probability distribution $\mu_\rho$ over the set of value maps that
accounts for the measured values of observables in \scr O.  The
precise meaning of ``accounts for'' is as follows.  For each
observable $A\in\scr O$, there is a probability distribution
$\mu_\rho^A$ induced on the spectrum of $A$ by
\[
\mu_\rho^A(X)=\mu_\rho(\{v:v(A)\in X\})
\]
for all subsets $X$ of the spectrum of $A$.  This induced probability
distribution should agree with the probability distribution predicted
by quantum theory for the observable $A$ in the state $\rho$.

One would thus expect that a no-go theorem in this context would say
that there is no way to assign, to each state, an appropriate
probability distribution over value maps.  Surprisingly, the no-go
theorems of Bell \cite{bell64,bell66} and Kochen and Specker \cite{ks}
are far stronger.  They say that, for \scr H of dimension at least 3,
there are no value maps at all for \scr H and the set $\scr
O_{\text{all}}$ of all self-adjoint operators on \scr H.  Better yet,
there are no value maps for certain specific finite\footnote{In the
  case of finite-dimensional \scr H, where each observable has only a
  finite spectrum, we can use the compactness theorem of propositional
  logic to infer, from the no-go theorem for $\scr O_{\text{all}}$,
  that there is also a no-go theorem for some finite $\scr
  O\subseteq\scr O_{\text{all}}$.  The compactness argument does not,
  however, produce a specific example of such an \scr O.} subsets \scr
O of $\scr O_{\text{all}}$.

We strengthen this result by tightly restricting the sort of
observables that are needed in \scr O.  This is
Theorem~\ref{val-thm-intro} from the introduction.

\begin{thm}     \label{val-thm}
Suppose that the dimension of the Hilbert space is at least 3.
\begin{lsnum}
\item There is a finite set \scr O of projections for which no value
  map exists.
\item If the dimension is finite then there is a finite set $\scr O$
  of rank~1 projections for which no value map exists.
\end{lsnum}
\end{thm}

The desired finite sets of projections are constructed explicitly in
the proof.

\begin{rmk} \label{findim} The assumption in part (2) of
  Theorem~\ref{val-thm} that the dimension of \scr H is finite cannot
  simply be omitted. If $\dim(\scr H)$ is infinite, then the set \scr
  O of all finite-rank projections admits a value map, namely the
  constant zero function.  This works because the definition of
  ``value map'' imposes constraints on only finitely many observables
  at a time.
\end{rmk}

\begin{proof}
  We start with proving Theorem~\ref{val-thm}.2, i.e. part (2) of
  Theorem~\ref{val-thm}.  Arguably the result is implicit in
  \cite[Section~5]{bell66} but it is not explicitly stated there and
  no specific \scr O of the desired sort is given.  In \cite{ks} and
  \cite{mermin}, the result is explicitly proved for 3-dimensional
  \scr H, but the extension to larger \scr H, which is easy if one
  just wants to extend a general no-go theorem, is not quite so
  obvious under the restriction to finitely many rank-1 projections.
  Because of this situation, we outline both versions of the proof,
  referring to these older papers for much of the work but filling in
  the additional arguments needed to get our result.

\begin{pf}[Proof of Theorem~\ref{val-thm}.2 following Bell]
  Bell \cite[Section~5]{bell66} works from three basic properties of
  (what we call) a value map $v$, namely
\begin{lsnum}
  \item For every rank-1 projection $\ket\psi\bra\psi$ (where
    \ket\psi\ is a unit vector), $v(\ket\psi\bra\psi)$ is 0 or 1.
\item If $v(\ket\phi\bra\phi)=1$ and \ket\psi\ is orthogonal to
  \ket\phi, then $v(\ket\psi\bra\psi)=0$.
\item If $v(\ket{\psi_1}\bra{\psi_1})=v(\ket{\psi_2}\bra{\psi_2})=0$
  for two orthogonal unit vectors $\ket{\psi_1}$ and $\ket{\psi_2}$,
  then also $v(\ket\psi\bra\psi)=0$ for all unit vectors \ket\psi\ of
  the form $\alpha\ket{\psi_1}+\beta\ket{\psi_2}$.
\end{lsnum}
All three of these follow from the definition of value map provided
\scr O contains all of the rank-1 projections of \scr H.  Property~(1)
is immediate from the fact that the spectrum of a non-trivial
projection is included in $\{0,1\}$.  Similarly, Property~(2) follows
from the facts that, if \ket\phi\ and \ket\psi\ are orthogonal, then
the projections $\ket\phi\bra\phi$ and $\ket\psi\bra\psi$ commute and
their joint spectrum is $\{(0,0),(0,1),(1,0)\}$.  (If \scr H were only
2-dimensional, this joint spectrum would be only $\{(0,1),(1,0)\}$,
but Property~(2) would still follow for the same reason: $(1,1)$ is
not in the joint spectrum.)

To prove Property~(3), complete $\{\ket{\psi_1},\ket{\psi_2}\}$ to an
orthonormal basis for \scr H, say
$\{\ket{\psi_1},\ket{\psi_2},\dots,\ket{\psi_n}\}$.  The associated
rank-1 projections $\ket{\psi_i}\bra{\psi_i}$ commute, and their joint
spectrum consists of the vectors in which one component is 1 and all
the rest are 0.  So we must have $v(\ket{\psi_i})=1$ for some
$i\geq2$.  But then the desired equation in (3) follows from (2)
because $\alpha\ket{\psi_1}+\beta\ket{\psi_2}$ is orthogonal to
$\ket{\psi_i}$. (This argument appears to require $\dim(\scr H)\geq 3$
in order to have a $\ket{\psi_i}$ to work with here, but this
appearance is wrong. If $\dim(\scr H)=2$ then Property~(3) holds
vacuously because $\{\ket{\psi_1},\ket{\psi_2}\}$ is an orthonormal
base for \scr H, so $v$ must send one of the associated projections to
1.  The real use of $\dim(\scr H)\geq 3$ comes later.)

Bell deduces from these three properties and $\dim(\scr H)\geq 3$ that
$v$ is continuous. More explicitly, he shows that, if
$v(\ket\phi\bra\phi)=0$ and $v(\ket\psi\bra\psi)=1$, for unit vectors
\ket\phi\ and \ket\psi, then $\nm{\ket\phi-\ket\psi}>\frac12$.  His
argument involves applying the three properties to some auxiliary
vectors in addition to \ket\phi\ and \ket\psi.  Bell completes the
proof of the no-go theorem by observing that, since $v$ must take both
values 0 and 1, this continuity result is a contradiction. So there
cannot be a value map defined on all of the rank-1 projections.

For our purposes, namely producing a finite set \scr O of rank-1
projections with no value map, we must work a bit more.  Using the
fact that $\dim(\scr H)$ is finite and at least 2, start with an
orthonormal base $\scr O_1$ for \scr H and enlarge it to a finite
superset $\scr O_2$ with the property that every two vectors
$\ket\phi,\ket\psi\in\scr O_2$ can be joined by a chain in $\scr O_2$,
\[
\ket\phi=\ket{\chi_0},\ket{\chi_1},\dots,\ket{\chi_l}=\ket\psi
\]
in which the distance between any two consecutive terms is at most
$\frac12$.  So, for each two consecutive terms, Bell's argument gives
us $v(\ket{\chi_i}\bra{\chi_i})=v(\ket{\chi_{i+1}}\bra{\chi_{i+1}})$.
Of course, the argument involves the auxiliary vectors mentioned
above, in addition to these two consecutive \ket\chi's, but there are
only finitely many of these auxiliary vectors.  Adjoin all of those
vectors, for all $i$, to $\scr O_2$ to get the final $\scr O$.  If $v$
were a value map for \scr O, then, by Bell's argument, we would have
$v$ constant on the rank-1 projections associated to the vectors in
$\scr O_2$ and therefore in particular the vectors in the orthonormal
base $\scr O_1$.  That is absurd, because a value map, when applied to
the projections associated to an orthonormal base always produces a
single 1 and the rest 0's.  So \scr O is as required by the theorem.
\end{pf}

\begin{pf}[Proof of Theorem~\ref{val-thm}.2 following Kochen-Specker and Mermin]
  When the dimension of \scr H is exactly 3, the constructions given
  by Kochen and Specker \cite{ks} and Mermin \cite[Section~IV]{mermin}
  provide the desired \scr O.  More precisely, the proof of Theorem~1
  in \cite{ks} uses a Boolean algebra generated by a finite set of
  one-dimensional subspaces of \scr H, and it shows that the
  projections to those subspaces constitute an \scr O of the required
  sort.  Mermin works instead with squares $S_i^2$ of certain
  spin-components of a spin-1 particle, but these are projections to
  2-dimensional subspaces of \scr H, and the complementary rank-1
  projections $I-S_i^2$ serve as the desired \scr O.

When the dimension of \scr H is greater than 3, but still finite, we
shall see in Theorem~\ref{reduce1} below how to bootstrap the result from lower to higher dimensions.  Notice that, if one merely wants a no-go
theorem saying that some \scr O has no value map, then this
bootstrapping is easy, as noted in \cite{bell64,ks,mermin}.  Work is
needed only to get all the operators in \scr O to be rank~1 projections.
\end{pf}

\begin{proof}[Proof Theorem~\ref{val-thm}.1]
The case where $\dim(\scr H)$ is finite was covered by
Theorem~\ref{val-thm}.2, so it remains to treat the case of
infinite-dimensional \scr H.

Let \scr K and \scr L be Hilbert spaces, with $\dim(\scr K)=3$ and
$\dim(\scr L)=\dim(\scr H)$.  Note that then their tensor product
$\scr K\otimes\scr L$ has the same dimension as \scr H, so it can be
identified with \scr H.

  Let \scr O be as in Theorem~\ref{val-thm} for the 3-dimensional \scr
  K.    Let
\[
\scr O'=\{P\otimes I_{\scr L}:P\in\scr O\},
\]
where $I_{\scr L}$ is the identity operator on \scr L.  Then $\scr O'$
is a set of infinite-rank projections of $\scr K\otimes\scr L=\scr
H$, having the same algebraic structure as \scr O.  It follows that
there is no value map for $\scr O'$.
\end{proof}

This completes the proof of Theorem~\ref{val-thm}.
\end{proof}

We note that the measurements involved in Theorem~\ref{val-thm}.2,
namely the rank-1 projections, are the same as those involved in our
expectation no-go Theorem~\ref{exp-thm}.  We hope that, by reducing
both species of no-go theorems to an extremely simple sort of
measurement, and furthermore a sort where measurement as observable
and measurement as effect coincide, we have clarified the similarities
as well as the differences between the two species.

\section{Bootstrapping the dimension}    \label{reduce}

Our objective in this section is to show that, in many cases, a no-go
theorem for a Hilbert space \scr H automatically yields no-go theorems
for larger Hilbert spaces, ones that contain \scr H as closed
subspaces.  The section has independent value and can be read
independently except that it needs the definition of value map and two
definitions (Spekkens's and ours) of probability representation.

Intuitively, such dimension bootstrapping results are to be expected.
If hidden-variable theories could explain the behavior of quantum
systems described by the larger Hilbert space, say $\scr H'$, then
they could also provide an explanation for systems described by the
subspace \scr H.  The latter systems are, after all, just a special
case of the former, consisting of the pure states that happen to lie
in \scr H or mixtures of such states.

The no-go theorems under discussion here, both ours
(Theorems~\ref{exp-thm} and \ref{val-thm}) and those from the previous
literature
(\cite{spekkens,ferrie2008,ferrie2009,ferrie2010,bell64,bell66,ks,mermin}),
give much more information than just the impossibility of matching the
predictions of quantum-mechanics with a hidden-variable theory.  They
establish that hidden-variable theories must fail in very specific
ways. It is not so obvious that these specific sorts of failures, once
established for a Hilbert space \scr H, necessarily also apply to its
superspaces $\scr H'$.

We shall prove two theorems saying that no-go results for a Hilbert
space $\scr H'$ follow directly from no-go results for a subspace
\scr H. The two theorems differ in the sort of no-go results that they
apply to; one is for expectation no-go results as in
Theorem~\ref{exp-thm}; the other is for value no-go results as in
Theorem~\ref{val-thm}.  We shall also comment on the situation for the
results in \cite{spekkens,ferrie2008,ferrie2009}.

We begin with the theorem dealing with value no-go results.  This is
the most important part of this section, because it was used in the
proof of Theorem~\ref{val-thm}.2 above. There, we invoked constructions
from the literature proving the result for \scr H of dimension 3 but
we claimed the result for all finite dimensions from 3 up.  That claim
is supported by the following theorem.

\begin{thm}             \label{reduce1}
Suppose $\scr H\subseteq\scr H'$ are finite-dimensional Hilbert
spaces.  Suppose further that \scr O is a finite set of rank-1
projections of \scr H for which no value map exists.  Then there is a
finite set $\scr O'$ of rank-1 projections of $\scr H'$ for which no
value map exists.
\end{thm}

\begin{pf}
  Clearly, if two Hilbert spaces are isomorphic and if one of them has
  a finite set \scr O of rank-1 projections with no value map, then
  the other also has such a set.  It suffices to conjugate the
  projections in \scr O by any isomorphism between the two
  spaces. Thus, the existence of such a set \scr O depends only on the
  dimension of the Hilbert space, not on the specific space.

  Proceeding by induction on the dimension of $\scr H'$, we see that
  it suffices to prove the theorem in the case where $\dim(\scr
  H')=\dim(\scr H)+1$.  Given such \scr H and $\scr H'$, let \ket\psi\
  be any unit vector in $\scr H'$, and observe that its orthogonal
  complement, $\ket\psi^\bot$, is a subspace of $\scr H'$ of the same
  dimension as \scr H and thus isomorphic to \scr H.  By the induction
  hypothesis, this subspace $\ket\psi^\bot$ has a finite set \scr O of
  rank-1 projections for which no value map exists.  Each element of
  \scr O can be regarded as a rank-1 projection of $\scr H'$; indeed,
  if the projection was given by $\ket\phi\bra\phi$ in
  $\ket\psi^\bot$, then we can just interpret the same formula
  $\ket\phi\bra\phi$ in $\scr H'$, using the same unit vector
  $\ket\phi\in\ket\psi^\bot$

  Let $\scr O_1$ consist of all the projections from \scr O,
  interpreted as projections of $\scr H'$, together with one
  additional rank-1 projection, namely $\ket{\psi}\bra{\psi}$.  What
  can a value map $v$ for $\scr O_1$ look like? It must send
  $\ket{\psi}\bra{\psi}$ to one of its eigenvalues, 0 or 1.

Suppose first that $v(\ket{\psi}\bra{\psi})=0$. Then, using the fact
that $\ket{\psi}\bra{\psi}$ commutes with all the other elements of
$\scr O_1$, we easily compute that what $v$ does to those other
elements amounts to a value map for \scr O.  But \scr O was chosen so
that it has no value map, and so we cannot have
$v(\ket{\psi}\bra{\psi})=0$.  Therefore $v(\ket{\psi}\bra{\psi})=1$.  (It
follows that $v$ maps the projections associated to all the other
elements of $\scr O'$ to zero, but we shall not need this fact.)

We have thus shown that any value map for the finite set $\scr O_1$
must send $\ket\psi\bra\psi$ to 1.  Repeat the argument for another
unit vector $\ket{\psi'}$ that is orthogonal to \ket\psi.  There is a
finite set $\scr O_2$ of rank-1 projections such that any value map
for $\scr O_2$ must send \ket{\psi'}\bra{\psi'} to 1.  No value map
can send both \ket\psi\bra\psi\ and \ket{\psi'}\bra{\psi'} to 1,
because their joint spectrum consists of only $(1,0)$ and $(0,1)$.
Therefore, there can be no value map for the union $\scr O_1\cup\scr
O_2$, which thus serves as the $\scr O'$ required by the theorem.
\end{pf}

The finiteness of $\dim(\scr H')$ is essential in this theorem.  If
the theorem were true for infinite-dimensional $\scr H'$, then the
same would be the case for Theorem~\ref{val-thm}, contrary to
Remark~\ref{findim}.  The next theorem, in contrast, does not require
dimensions to be finite.

\begin{thm}             \label{exp-up}
Let $\scr H'$ be a Hilbert space and $\scr H$ a closed subspace of
$\scr H'$.  From any probability representation (our version) for
quantum systems described by $\scr H'$, one can directly construct
such a representation for systems described by $\scr H$.
\end{thm}

Strictly speaking, this theorem is vacuous, since
Theorem~\ref{exp-thm} says that there is no probability representation
(our version) for quantum systems described by any Hilbert space of
dimension $\geq2$.  The intention, however, is that the construction
here is considerably easier than that in Theorem~\ref{exp-thm}.  In
particular, if we knew  Theorem~\ref{exp-thm} only for 2-dimensional
\scr H, this would suffice to get the full  Theorem~\ref{exp-thm}.
This fact supports our assessment, in Section~\ref{exp}, that the
careful development and rigorous proofs in \cite{ferrie2010} are a
greater contribution than the extension to infinite-dimensional
Hilbert spaces.  (Additional support will come later in this section.)

\begin{pf}
  We construct a probability representation (our version) $\Lambda$,
  $T$, and $S$ for quantum systems described by \scr H (with notation
  as in Definition~\ref{v3}) from any such representation $\Lambda'$,
  $T'$, and $S'$ for the larger Hilbert space $\scr H'$.  To begin, we
  set $\Lambda=\Lambda'$.

  To define $T$ and $S$, we use the inclusion map $i:\scr H\to\scr
  H'$, sending each element of \scr H to itself considered as an
  element of $\scr H'$, and we use the adjoint $p:\scr H'\to\scr H$,
  which is the orthogonal projection of $\scr H'$ onto \scr H.  Given
  any density operator $\rho\in\scr T_{+1}(\scr H)$, we can expand it
  to a density operator $\bar\rho=i\circ\rho\circ p\in\scr T_{+1}(\scr
  H')$.  Note that this expansion is very natural: If $\rho$
  corresponds to a pure state $\ket\psi\in\scr H$, i.e., if
  $\rho=\ket\psi\bra\psi$, then $\bar\rho$ corresponds to the same
  $\ket\psi\in\scr H'$.  If, on the other hand, $\rho$ is a mixture of
  states $\rho_i$, then $\bar\rho$ is the mixture, with the same
  coefficients, of the $\overline{\rho_i}$.  Define $T:\scr T_{+1}(\scr
  H)\to\scr M_{+1}(\Lambda)$ by $T(\rho)=T'(\bar\rho)$.

The definition of $S$ is similar. Notice that, if $E$ is a rank-1
projection in \scr H, then $\bar E=i\circ E\circ p$ is a rank-1
projection in $\scr H'$.  So we can define $S(E)=S'(\bar E)$. Again,
the passage from $E$ to $\bar E$ is very natural.  If $E$ projects to
the one-dimensional subspace spanned by $\ket\psi\in\scr H$, then
$\bar E$ projects to the same subspace, now considered as a subspace
of $\scr H'$.

This completes the definition of $\Lambda$, $T$ , and $S$.  Most of
the requirements in Definition~\ref{v3} are trivial to verify.  For
the last requirement, the agreement between the expectation computed
as a trace in quantum mechanics and the expectation computed as an
integral in the probability representation, it is useful to notice
first $p\circ i$ is the identity operator on $\scr H$.  We can then
compute, for any $\rho\in\scr T_{+1}(\scr H)$ and any rank-1
projection $E$ on \scr H,
\begin{align*}
  \int_\Lambda S(E)\,dT(\rho)&=\int_\Lambda S'(\bar E)\,dT'(\bar\rho) \\
  &=\tr{\bar\rho\bar E}\\
  &=\tr{i\circ\rho\circ p\circ i\circ E\circ p}\\
  &=\tr{i\circ\rho\circ E\circ p}\\
  &=\tr{\rho\circ E\circ p\circ i}\\
  &=\tr{\rho\circ E},
\end{align*}
as required.
\end{pf}

To finish this section, we briefly discuss the possibility of
transferring no-go theorems as in
\cite{spekkens,ferrie2008,ferrie2009}   from a Hilbert space \scr H to a
larger space $\scr H'$.  To be specific, we consider probability
representations (Spekkens version) as in Definition~\ref{v1}, subject to the
assumptions of determinateness ($\xi_E$ depends only on the effect
$E$, not on the POVM containing it) and convex-linearity of both of
the maps $\rho\mapsto\mu_\rho$ and $E\mapsto\xi_E$.

\begin{prop}            \label{reduce-spek}
Let $\scr H$ be a closed subspace of the Hilbert space $\scr H'$.  If
$\scr H'$ admist a probability representation (Spekkens version)
satisfying determinateness and convex-linearity, then so does \scr H.
\end{prop}

\begin{pf}
  At first, it might seem that we can proceed exactly as in the proof
  of Theorem~\ref{exp-up}, transforming the density operators $\rho$
  and effects $E$ of the subspace \scr H to density operators
  $\bar\rho=i\circ\rho\circ p$ and effects $\bar E=i\circ E\circ p$ on
  the superspace $\scr H'$, and then using this transformation to
  convert a probability representation (Spekkens version) for $\scr
  H'$, say $\Lambda',\mu',\xi'$, to one for \scr H.  In detail, we
  would use the same measure space, $\Lambda=\Lambda'$, and we would
  set $\mu_\rho=\mu'_{\bar\rho}$ and $\xi_E=\xi'_{\bar E}$.

This approach works well as far as $\bar\rho$ and $\mu_\rho$ are
concerned, but there is a problem with $\bar E$ and $\xi_E$.
Definition~\ref{v1} requires that, if $\{E_k:k\in K\}$ is a POVM, i.e.,
if the effects $E_k$ have sum $I$, then $\sum_{k\in
  K}\xi_{E_k}(\lambda)=1$ for all $\lambda\in\Lambda$.  Given that
$\xi'$ satisfies this requirement on $\scr H'$, we want that $\xi$
satisfies it on \scr H.  So we would like to argue that, if
$\{E_k:k\in K\}$ is a POVM in \scr H, then $\{\bar E_k:k\in K\}$ is a
POVM in $\scr H'$, which would give us that
$\sum_k\xi_{E_k}(\lambda)=\sum_k\xi'_{\bar E_k}(\lambda)=1$.
Unfortunately, $\{\bar E_k:k\in K\}$ will not be a POVM for $\scr H'$
(unless $\scr H=\scr H'$).  Indeed, using the fact that $\{E_k:k\in
K\}$ is a POVM, we can compute
\[
\sum_k\bar E_k=\sum_ki\circ E_k\circ p=i\circ\Big(\sum_{k\in
  K}E_k\Big)\circ p=i\circ I\circ p=i\circ p.
\]
Here $i\circ p$ is the transformation $i\circ p:\scr H'\to\scr H'$
that projects orthogonally to the subspace \scr H; it is not the
identity unless $\scr H=\scr H'$.

To correct the problem, we modify the definition of $\bar E$ as
follows.  Fix an arbitrary unit vector $\ket\alpha\in\scr H$.  Then
define $\bar E$ to be the unique linear operator on $\scr H'$ such that
\[
\bar E\ket\psi =
\begin{cases}
E\ket\psi & \text{if }\psi\in\scr H,\\
\bra\alpha E\ket\alpha\ket\psi & \text{if }\ket\psi\bot\scr H.
\end{cases}
\]
In other words, $\bar E$ agrees with $E$ on $\scr H$ and with a scalar
multiple of the identity on the orthogonal complement of $\scr H$, the
multiplier of the identity being $\bra\alpha E\ket\alpha$.  Another
way to write $\bar E$ uses the operator $I-i\circ p$, which projects
$\scr H'$ onto the orthogonal complement of \scr H; we have
\[
\bar E=i\circ E\circ p+\bra\alpha E\ket\alpha(I-i\circ p).
\]
This new version of $\bar E$ overcomes the problem with the old one,
because, if $\sum_kE_k=I$, then, because \ket\alpha\ is a unit vector,
$\sum_k \bra\alpha E_k\ket\alpha=1$ and
\[
\sum_k\bar E_k=\sum_ki\circ E_k\circ p+\sum_k\bra\alpha E_k\ket\alpha
(I-i\circ p) =i\circ p+1(I-i\circ p)=I.
\]

Furthermore, this extension process from $E$ to $\bar E$ sends the
identity and zero operators on $\scr H$ to the identity and zero
operators on $\scr H'$, and the process respects weighted averages.
Using the new extension process, we define $\xi_E=\xi'_{\bar E}$, and
we claim that the result is a probability representation (Spekkens version)
for \scr H.  The only non-trivial thing to check is the final
requirement that the quantum-theoretic expectation values $\tr{\rho
  E}$ agree with the hidden-variable theory's expectation values $\int
d\lambda\,\mu_{\rho}(\lambda)\xi_E(\lambda)$.  We compute
\begin{align*}
\int_\Lambda d\lambda\,\mu_{\rho}(\lambda)\xi_E(\lambda)&=
\int_\Lambda d\lambda\,\mu'_{\bar\rho}(\lambda)\xi'_{\bar
  E}(\lambda)\\
&=\tr{\bar\rho\bar E}\\
&=\tr{i\rho p\cdot(iE\rho+\bra\alpha E\ket\alpha(I-ip))}\\
&=\tr{i\rho piEp}+\bra\alpha E\ket\alpha\tr{i\rho p(I-ip)}.
\end{align*}
The first term here was computed earlier and found to be \tr{\rho E},
which is the desired result, so it remains to check that the second
term vanishes.  Up to a factor $\bra\alpha E\ket\alpha$, it is
\[
\tr{i\rho p-i\rho pip}=\tr{i\rho p-i\rho p}=0,
\]
where we have used that $pi$ is the identity operator of \scr H.  This
completes the proof of the proposition.
\end{pf}

Thus, for example, to prove the no-go theorems of Spekkens
\cite{spekkens} and of Ferrie and Emerson \cite{ferrie2008,ferrie2009}
(with appropriate clarifications as discussed above in
Section~\ref{exp}), it would suffice to prove them for two-dimensional
Hilbert spaces (in quantum computing terminology, one-qubit spaces);
the theorems would automatically carry over to all larger Hilbert
spaces.  Because of the need for clarifications in these theorems, we
give, in Appendix~\ref{old}, a proof of a Spekkens-style no-go theorem
for Hilbert spaces of dimension two.

\begin{rmk}
The proof of Proposition~\ref{reduce-spek} involved choosing an
arbitrary unit vector \ket\alpha\ in \scr H.  This arbitrariness can
be avoided when \scr H is finite-dimensional by averaging over all
\ket\alpha's.  That is, if $\dim(\scr H)=d$, then we can replace the
definition of $\bar E$ in the proof with
\[
\bar E\ket\psi=
\begin{cases}
E\ket\psi & \text{if }\psi\in\scr H,\\
\frac1d\tr E\ket\psi & \text{if }\ket\psi\bot\scr H,
\end{cases}
\]
and, since $\tr I=d$, the rest of the proof would work as before.
\end{rmk}

\section{Bell's Example and Symmetry}        \label{example}

Theorem~\ref{exp-thm} applies to all Hilbert spaces of dimension at
least~2.  We cannot expect any sort of no-go result in lower
dimensions, because quantum theory in Hilbert spaces of dimensions 0
and 1 is trivial and therefore classical.  The second part of
Theorem~\ref{val-thm} applies only to Hilbert spaces whose dimension
is finite and at least 3.  We have already indicated in
Remark~\ref{findim} why the theorem fails in infinite dimensions and
in the first part of Theorem~\ref{val-thm} why a modified version
holds in infinite dimensions.  What about dimension 2?

Bell has given, in \cite{bell64,bell66}, hidden-variable theories for
a two-dimensional Hilbert space.  More precisely, he has assigned to
each pure state $\ket\psi$ in such a Hilbert space \scr H a
probability distribution on value maps, such that the resulting
probability distributions for any observable agree with the
predictions of quantum theory.  In this section, we summarize the
improved version of Bell's example described by Mermin \cite{mermin},
we simplify part of his argument, and we explain why the example
doesn't contradict Theorem~\ref{exp-thm}.

We work with the Hilbert space \scr H of 2-component vectors over \bbb
C, so that operators on \scr H are given by $2\times2$ matrices.  Let
$\vec\sigma$ be the 3-component ``vector'' whose entries are the Pauli
matrices
\[
\sigma_x=
\begin{pmatrix}
  0&1\\1&0
\end{pmatrix},\qquad \sigma_y=
\begin{pmatrix}
  0&-i\\i&0
\end{pmatrix},\qquad \sigma_z=
\begin{pmatrix}
  1&0\\0&-1
\end{pmatrix}.
\]
If $\vec n$ is any 3-component unit vector in $\bbb R^3$, then the dot
product $\vec n\cdot\vec\sigma$ is a Hermitian operator with
eigenvalues $\pm1$.  Every pure state of \scr H is an eigenstate, for
eigenvalue $+1$, of $\vec n\cdot\vec\sigma$ for a unique $\vec n$.  We
use the notation \ket{\vec n} for this eigenstate.  (If \scr H
represents the states of a spin-$\frac12$ particle, then the operator
$\frac12\vec n\cdot\vec\sigma$ represents the spin component in the
direction $\vec n$, and so \ket{\vec n} represents the state in which
the spin is definitely aligned in the direction $\vec n$.  It is a
special property of spin $\frac12$ that all pure states are of this
form; for higher spins, a superposition of states with definite spin
directions need not have a definite spin direction.)

Any observable, i.e., any Hermitian operator on \scr H, can be
expressed as $A=a_0I+(\vec a\cdot\vec\sigma)$ for some scalar
$a_0\in\bbb R$ and vector $\vec a\in\bbb R^3$.  Its eigenvalues are
$a_0\pm\nm{\vec a}$.

The hidden-variable theory, as presented in \cite[Section~3]{mermin},
assigns to each state $\ket{\vec n}$ a family of sub-ensembles labeled
by unit vectors $\vec m\in\bbb R^3$, the probability distribution of
$\vec m$ being uniform on the unit sphere in $\bbb R^3$.  In the
sub-ensemble of $\ket{\vec n}$ given by $\vec m$, the observable
$a_0I+(\vec a\cdot\vec\sigma)$ has the (definite) value
\begin{align*}
  a_0+\nm{\vec a}&\quad\text{if }(\vec m+\vec n)\cdot\vec a\geq0\\
a_0-\nm{\vec a}&\quad\text{if }(\vec m+\vec n)\cdot\vec a<0.
\end{align*}
Mermin writes that elementary integration confirms that, for any fixed
state \ket{\vec n}, the average over all $\vec m$ of the values
asigned to an observable $a_0I+(\vec a\cdot\vec\sigma)$ agrees with
the result $a_0+(\vec a\cdot\vec n)$ predicted by quantum mechanics.
In fact, the required integration is so elementary that it was done by
Archimedes.  All one needs is the theorem that, when a sphere is cut
by a plane, its area is divided in the same ratio as the length of the
diameter perpendicular to the plane.  To verify that the average over
$\vec m$ of the values of $a_0I+(\vec a\cdot\vec\sigma)$ in the state
\ket{\vec n} is $a_0+(\vec a\cdot\vec n)$, we begin with a couple of
simplifications.  First, we may assume that $a_0=0$, because a general
$a_0$ would just be added to both sides of the equation that we are
trying to prove.  Second, thanks to the rotational symmetry of the
situation (where any rotation is applied to all three of $\vec a$,
$\vec n$ and $\vec m$), we may assume that the vector $\vec a$ points
in the $z$-direction.  Finally, by scaling, we may assume that $\vec
a=(0,0,1)$.  So our task is to prove that the average over $\vec m$ of
the values assigned to $\sigma_z$ is $n_z$.  By definition, the value
assigned to $\sigma_z$ is $\pm1$, where the sign is chosen to agree
with that of $m_z+n_z$.  In view of how $\vec m$ is chosen, this
$m_z+n_z$ is the $z$-coordinate of a random point on the unit sphere
centered at $\vec n$.  So the question reduces to determining what
fraction of this sphere lies above the $x$-$y$ plane.  This plane cuts
this unit sphere horizontally at a level $n_z$ below the sphere's
center.  So, by Archimedes's theorem, it divides the sphere's area in
the ratio of $1+n_z$ (above the plane) to $1-n_z$ (below the plane).
That is, the value assigned to $\sigma_z$ is $+1$ with probability
$(1+n_z)/2$ and $-1$ with probability $(1-n_z)/2$.  Thus, the average
value of $\sigma_z$ is $n_z$, as required.

This hidden-variable theory can be viewed in the framework of
Section~\ref{val}.  Each of the vectors $\vec m+\vec n$ corresponds to
a value map, namely the map sending any observable $a_0I+(\vec
a\cdot\vec\sigma)$ to the value described above.  It is not difficult
to verify that this is indeed a value map, because there are so few
commuting observables for our 2-dimensional \scr H.  Two observables
commute if and only if their $\vec a$'s are parallel or antiparallel.
That is, they differ by only a scalar factor on the $\vec
a\cdot\vec\sigma$ part and an arbitrary change of the $a_0I$ part.

The mere existence of a value map (let alone a good probabiilty
distribution on value maps for all the states) shows that, in
Theorem~\ref{val-thm}, the hypothesis of dimension $\geq3$ cannot be
weakened so as to allow dimension 2.

What happens if we try to fit this hidden-variable theory into the
framework of Section~\ref{exp}?  A natural choice for $\Lambda$ is the
space of all the value maps obtained above, or, more geometrically,
the space of their parametrizations $\vec m+\vec n$.  Since both $\vec
m$ and $\vec n$ are unit vectors, $\Lambda$ will be the ball of radius
2 centered at the origin of $\bbb R^3$.  For any pure state $\ket{\vec
  n}$, the associated probability distribution $T(\ket{\vec
  v}\bra{\vec v})$ is the uniform distribution on the two-dimensional
surface of a unit sphere centered at $\vec n$, because we are choosing
$\vec m$ randomly while $\vec n$ is fixed.  Notice that the framework
of Definition~\ref{v1} does not handle this situation well, because
these probability distributions are not absolutely continuous with
respect to any natural probability distribution on $\Lambda$.  (What a
physicist might call the probability density on $\Lambda$ associated
to a state is not a function but a distribution.) So we work instead
with the framework of Ferrie, Morris, and Emerson \cite{ferrie2010},
as summarized in Definition~\ref{v2} above or with the more liberal
Definition~\ref{v3}.

Both of these definitions require a convex-linear map $T$ from the set
$\scr T_{+1}$ of density matrices (representing mixed states) to the
set $\scr M_{+1}$ of probability measures on $\Lambda$.  The
hidden-variable theory under consideration has, so far, provided
measures only for the pure states, i.e., the density matrices of the
special form $\ket{\vec n}\bra{\vec n}$; to such a density matrix, it
associated the uniform measure on the unit sphere surface centered at
$\vec n$.  To obtain a probability representation, in either the
Ferrie-Morris-Emerson version or our version, we must extend this map
convex-linearly to all density matrices.

No such extension exists.  Here is an example showing what goes wrong.
Consider the four pure states corresponding to spin in the directions
of the positive $x$, negative $x$, positive $z$ and negative $z$
axes.  The corresponding density operators are the projections
\[
\frac{I+\sigma_x}2,\quad\frac{I-\sigma_x}2,\quad
\frac{I+\sigma_z}2,\quad \frac{I-\sigma_z}2,
\]
respectively.  Averaging the first two with equal weights, we get
$\frac12 I$; averaging the last two gives the same result.  So a
convex-linear extension $T$ would have to assign to the density
operator $\frac12I$ the average of the probability measures assigned
to the pure states with spins in the $\pm x$ directions and also the
average of the probability measures assigned to pure states with spins
in the $\pm z$ directions.  But these two averages are visibly very
different.  The first is concentrated on the union of two unit spheres
tangent to the $y$-$z$-plane at the origin, while the second is
concentrated on the union of two unit spheres tangent to the
$x$-$y$-plane at the origin.

Thus, Bell's example of a hidden-variable theory for 2-dimensional
\scr H does not fit the assumptions in any of the expectation no-go
theorems.  It does not, therefore, clash with the fact that those
theorems, unlike the value no-go theorems, apply in the 2-dimensional
case.

Another way to view this situation is as a demonstration that the
hypothesis of convex-linearity cannot be omitted from the expectation
no-go theorems.  In comparison with Definition~\ref{v2}, which
described the hypotheses used by Ferrie, Morris, and Emerson
\cite{ferrie2010}, our Definition~\ref{v3} dropped the requirement of
convex-linearity for effects; Bell's example shows that we cannot also
drop that requirement for states.

In view of the idea of symmetry or even-handedness suggested by
Spekkens \cite{spekkens}, one might ask whether there is a dual
version of Theorem~\ref{exp-thm}, that is, a version that requires
convex-linearity for effects but looks only at pure states and does not
require any convex-linearity for states.

The answer is no; with such requirements there is a trivial example of
a successful hidden-variable theory, regardless of the dimension of
the Hilbert space, so there cannot be a no-go theorem.  The example
can be concisely described as taking the quantum state itself as the
``hidden'' variable.  In more detail, let $\Lambda$ be the set of all
states, i.e., the projective space obtained from the set of unit
vectors of \scr H by identifying any two that differ only by a phase
factor.  Let $T$ assign to each pure state $\ket\psi\bra\psi$ the
probability measure on $\Lambda$ concentrated at the point
$\lambda_{\ket\psi}$ that
corresponds to the vector $\ket\psi$.  Let $S$ assign to each effect
$E$ the function on $\Lambda$ defined by
\[
S(E)(\lambda_{\ket\psi})=\bra\psi E\ket\psi.
\]
We have trivially arranged for this to give the correct expectation
for any effect $E$ and any pure state \ket\psi.  The formula for
$S(E)$ is clearly convex-linear (in fact, linear) as a function of
$E$.  Of course, $T$ cannot be extended convex-linearly to mixed
states, so that Theorem~\ref{exp-thm} does not apply.

\appendix
\section{Convex-Linearity}      \label{conv-lin}

As we pointed out, near the end of Section~\ref{spek:subsec}, Spekkens
\cite{spekkens} erroneously claims that, if a function $f$ is
convex-linear on a convex set \scr S of operators that span the space
of Hermitian operators (and $f$ takes the value zero on the zero
operator if the latter is in \scr S), then $f$ can be uniquely
extended to a linear function on this space.

The correct version of the result extends $f$ not to a linear function
but to translated-linear function, i.e., a composition of translations
and a linear function.  The rest of this section is devoted to a proof
of this fact, in its natural level of generality.  It applies to
arbitrary real vector spaces; that the space consists of Hermitian
operators is irrelevant.

The \emph{convex hull}, \cv S, of a subset $S$ of a real vector space
$V$ consists of the convex combinations $a_1v_1+\cdots+a_nv_n$ of
vectors $v_1,\dots,v_n\in S$ where $a_1+\cdots+a_n=1$ and every
$a_i\geq0$.  The \emph{affine hull}, \af S, of $S$ consists of the
affine combinations $a_1v_1+\cdots+a_nv_n$ of vectors
$v_1,\dots,v_n\in S$ where $a_1+\cdots+a_n=1$ but some coefficients
$a_i$ may be negative.

A set is \emph{convex} if it contains all the convex combinations of
its members; similarly, it is an \emph{affine} space if it contains
all the affine combinations of its members.  An easy computation shows
that convex hulls are convex and affine hulls are affine spaces; that
is $\cv{\cv S}=\cv S$ and $\af{\af S}=\af S$.

An affine space $A$ in a vector space $V$ is said to be \emph{parallel}
to a linear subspace $L$ of $V$ if $A=u_0+L=\{u_0+v:v\in L\}$ for some
$u_0\in V$.
It is easy to see that, if an affine space $A$ is
parallel to a linear space $L$ as above, then (i)~$L$ is unique,
(ii)~$u_0\in A$, (iii)~any vector in $A$ can play the role of the
translator $u_0$, and (iv)~$A$ is either equal to $L$ or disjoint from
$L$.

\begin{la}[\S1 in \cite{Rockafellar}]\label{Rockafellar}
Any affine subspace $A$ of a real vector space $V$ is parallel to a
linear subspace $L$ of $V$.
\end{la}

In other words, any affine subspace is a translation of a linear
subspace.  For example, in $\bbb R^2$, we have that
$\mathrm{Aff}\{(0,1),(1,0)\}$ is parallel to the diagonal $y=-x$, and
$\mathrm{Aff}\{(0,1),(1,0),(1,1)\}$ is (and thus is parallel to) $\bbb
R^2$.

\begin{proof}
If $A$ contains the zero vector $\vec0$ then it is a linear
subspace. Indeed, if $v\in A$ then any multiple $av = av +
(1-a)\vec0 \in A$. And if $u,v\in A$ then $u+v = 2(\frac12u +
\frac12v)\in A$.

For the general case, let $u_0$ be any vector in the affine space $A$.
It suffices to show that $L = \{v-u_0: v\in A\}$ is an affine space,
because then the preceding paragraph shows that it is a linear space,
and clearly $A=u_0+L$. Any affine combination
$a_1(v_1-u_0)+\cdots+a_n(v_n-u_0)$ of vectors in $L$ (so the $v_i$ are
in $A$ and the sum of the $a_i$ is 1) can be rewritten as
$(a_1v_1+\cdots+a_nv_n)-u_0$, which is in $L$.
\end{proof}

Let $V$ and $W$ be real vector spaces, $S$ a subset of $V$, $C=\cv S$
its convex hull, and $A=\af S$ its affine hull.  Recall that a
transformation $f:C\to W$ is convex-linear on $S$ if
\[
  f(a_1v_1 +\cdots+ a_nv_n) = a_1f(v_1) +\cdots+ a_nf(v_n)
\]
for any convex combination $a_1v_1 +\cdots+ a_nv_n$ of vectors $v_i$
from $S$.  A transformation $f:A\to W$ is \emph{translated-linear} if
it has the form $f(v)=w_0+h(v-u_0)$ for some $w_0\in W$, some $u_0\in
A$, and some linear function $h:L\to W$ defined on the linear space
$L=A-u_0$ parallel to $A$.

\begin{prop} \label{translinear} With notation as above, any
  transformation $f:C\to W$ that is convex-linear on $S$ has a unique
  extension to a translated-linear function on $A$.
\end{prop}

\begin{pf}
Notice first that translations $v\mapsto v-u_0$ and linear functions both
preserve affine combinations.  A translated-linear function, being the
composition of two translations and a linear function, therefore also
preserves affine combinations.  This observation implies the
uniqueness part of the proposition.  Indeed, every element of $A$ is
an affine combination $a_1s_1+\cdots+a_ns_n$ of elements of $S$, and
therefore any translated-linear extension of $f$ must map it to
$a_1f(s_1)+\cdots+a_nf(s_n)$.

To prove the existence part of the proposition, it will be useful to
work with the graphs of functions.  For any function $g:S\to W$ with
$S\subseteq V$, its \emph{graph} is the subset of $V\oplus W$
consisting of the pairs $(s,g(s))$ for $s\in S$.\footnote{In
  set-theoretic foundations, a function is usually defined as a set of
  ordered pairs, and so $g$ is the same thing as its graph.}  We
record for future reference that the graph of $g$ is a linear subspace
of $V\oplus W$ if and only if the domain of $g$ is a linear subspace
of $V$ and $g$ is a linear transformation from that domain to $W$.  We
also note that the projection $\pi:V\oplus W\to V:(v,w)\mapsto v$ is a
linear transformation that sends the graph of any $g$ to the domain of
$g$.

In the situation of the proposition, let $f:C\to W$ be a
transformation that is convex-linear on $S$, and let $F\subseteq
V\oplus W$ be its graph.  Also, let $F^-$ be the graph of the
restriction of $f$ to $S$.  Notice that the convex-linearity of $f$ on
$S$ means exactly that $F$ is the convex hull of $F^-$.  It follows
that $F$ and $F^-$ have the same affine hull, because
\[
\af F=\af{\cv{F^-}}\subseteq\af{\af{F^-}}=\af{F^-}\subseteq\af F.
\]
We claim that this affine hull $\af{F^-}$ is the graph of a function;
that is, it does not contain two distinct elements $(v,w)$ and
$(v,w')$ with the same first component $v$.  To see this, suppose we
had two such elements in $\af F=\af{F^-}$, say
\[
(v,w)=a_1(s_1,f(s_1))+\cdots+a_m(s_m,f(s_m))
\]
and
\[
(v,w')=b_1(t_1,f(t_1))+\cdots+b_n(t_n,f(t_n)),
\]
where all the $s_i$'s and $t_j$'s are in $S$ and where
\begin{equation}\label{rep1}
     a_1 +\cdots+ a_m = b_1 +\cdots+ b_n,
\end{equation}
because both sides are equal to 1. So we have
\begin{equation}\label{rep2}
    a_1s_1 +\cdots+ a_ms_m = b_1t_1 +\cdots+ b_nt_n ,
\end{equation}
because both sides are equal to $v$, and we want to prove $w=w'$,
i.e.,
\begin{equation}\label{rep3}
    a_1f(s_1) +\cdots+ a_mf(s_m) = b_1f(t_1) +\cdots+ b_nf(t_n ).
\end{equation}
In the special case where all coefficients $a_i$ and $b_j$ are $\ge0$,
vector $v$ is in $C$ and both sides of \eqref{rep3} are equal to
$f(v)$. The general case reduces to this special case as follows. In
all three equations \eqref{rep1}--\eqref{rep3}, move every summand
with a negative coefficient to the other side, and then divide the
resulting equations by the left part of the rearranged equation
\eqref{rep1}. As a result we return to the special case already
treated. Since the old version of \eqref{rep3} follows from the new
one, this completes the proof of our claim that $\af F=\af{F^-}$ is
the graph of a function.

By Lemma~\ref{Rockafellar}, the affine space $\af F$ is parallel to a
linear subspace $H$ of $V\oplus W$, say $\af F=(u_0,w_0)+H$, where
$u_0\in V$ and $w_0\in W$.  From the fact that \af F is the graph of a
function, it follows immediately that $H$ is also the graph of a
function.  Indeed, if $H$ contains $(v,w)$ and $(v,w')$, then \af F
contains $(v-u_0,w-w_0)$ and $(v-u_0,w'-w_0)$, so $w-w_0=w'-w_0$ and
$w=w'$.

Let $h$ be the function whose graph is $H$.  Because $H$ is a linear
subspace of $V\oplus W$, we know that $h$ is a linear transformation
from some linear subspace $L$ of $V$ into $W$.

The fact that $(u_0,w_0)+H=\af F$ tells us, by applying the linear projection
$\pi:V\oplus W\to V$, that $u_0+L$ equals
\[
\pi(\af F)=\af{\pi(F)}=\af C=A,
\]
where the first equality comes from linearity of $\pi$ and the second
from the fact that $F$ is the graph of the function $f$ whose domain
is $C$.  So $A$ is parallel to the linear subspace $L$ of $V$.
Furthermore, for each $v\in C$, we have
\[
(v,f(v))\in F\subseteq \af F=(u_0,w_0)+H,
\]
so $(v-u_0,f(v)-w_0)$ is in the graph $H$ of $h$.  That is,
$h(v-u_0)=f(v)-w_0$ and so $f(v)=w_0+h(v-u_0)$.  Thus, the
translated-linear function $v\mapsto w_0+h(v-u_0)$ is the desired
extension of $f$.
\end{pf}

\begin{rmk}
A linear function $h$ on a subspace $L$ of a vector space $V$ can be
extended to a linear function $\bar h$ on all of $V$.  Extend any
basis of $L$ to a basis of $V$, define $\bar h$ arbitrarily on the
new basis vectors that are not in $L$, and extend the resulting
function by linearity to all of $V$.

For transformations defined on
all of $V$, we have a simpler formula for translated-linear functions,
because
\[
w_0+\bar h(v-u_0)=w_0+\bar h(v)-\bar h(u_0)=\bar h(v)+w_1,
\]
where $w_1=w_0-\bar h(u_0)$.

On the other hand, in contrast to Proposition~\ref{translinear}, this
$\bar h$ is not unique (unless $L=V$).

Also, in the case of infinite-dimensional spaces, the extension
process requires the axiom of choice (to extend bases) and need not be
well-behaved with respect to natural topologies on the vector spaces.
\end{rmk}

\section{No-Go Theorem for Spekkens Version}       \label{old}

This appendix is devoted to proving the following no-go theorem for
the original Spekkens version of probability representations, subject
to the clarifications discussed in Section~\ref{spek:subsec}.

\begin{thm}     \label{no-go}
  For a Hilbert space \scr H of dimension at least two, there is no
probability representation (Spekkens version) subject to
determinateness and convex-linearity.
\end{thm}

\begin{pf}
In view of Proposition~\ref{reduce-spek}, it suffices to prove the
theorem under the assumption that \scr H has dimension exactly two.

To begin, we recall the form of density operators and effects in a
two-dimensional Hilbert space \scr H.  A basis for the Hermitian
operators on \scr H is given by the identity and the three Pauli
matrices
\[
I=
\begin{pmatrix}
  1&0\\0&1
\end{pmatrix},\quad \sigma_x=
\begin{pmatrix}
  0&1\\1&0
\end{pmatrix},\quad \sigma_y=
\begin{pmatrix}
  0&-i\\i&0
\end{pmatrix},\quad \sigma_z=
\begin{pmatrix}
  1&0\\0&-1
\end{pmatrix}.
\]
It will be convenient to use vector notation, denoting the triple  of
matrices $(\sigma_x,\sigma_y,\sigma_z)$ by $\vec \sigma$.  Then the
general Hermitian matrix
looks like
\[
wI+x\sigma_x+y\sigma_y+z\sigma_z=wI+\vec x\cdot\vec\sigma,
\]
where $w$ and the three components of $\vec x$ are real numbers.
The eigenvalues of this Hermitian matrix are
\[
w\pm\sqrt{x^2+y^2+z^2}=w\pm\nm{\vec x}
\]
In particular, the trace of this matrix is $2w$, and the matrix is
positive if and only if $w\geq\nm{\vec x}$.

Density matrices are the Hermitian, positive matrices of trace 1, so
they have the form
\[
\rho=\rho(\vec x)=\frac12(I+\vec x\cdot\vec\sigma),
\]
where $\nm{\vec x}\leq 1$.  As indicated by the notation, we
parametrize these density matrices by three-component vectors $\vec x$
of norm $\leq 1$.  The three-dimensional ball that serves as the
parameter space here is called the Bloch sphere (with its interior).

Similarly, effects have the form
\[
E=E(m,\vec p)=mI+p\sigma_x+q\sigma_y+r\sigma_z =
mI+\vec p\cdot\vec\sigma
\]
with
\[
\nm{\vec p}\leq m\leq 1-\nm{\vec p}
\]
(because $E$ and $I-E$ are positive operators) and therefore $\nm{\vec
  p}\leq\frac12$.    The parameter space here,
consisting of all four-component vectors satisfying these
inequalities, is a double cone over a three-dimensional ball of radius
$\frac12$.

We record for future reference the traces
\[
\text{Tr}(I)=2,\quad\text{Tr}(\sigma_x)=\text{Tr}(\sigma_y)
=\text{Tr}(\sigma_z)=0
\]
and the multiplication table
\[
\sigma_x\sigma_y=-\sigma_y\sigma_x=i\sigma_z, \quad
\sigma_y\sigma_z=-\sigma_z\sigma_y=i\sigma_x,\quad
\sigma_z\sigma_x=-\sigma_x\sigma_z=i\sigma_y,
\]
and
\[
 \sigma_x^2=\sigma_y^2=\sigma_z^2=I.
\]
From these facts, it is easy to compute that
\[
\text{Tr}(\rho(\vec x)E(m,\vec p))=m+\vec x\cdot\vec p,
\]
where the factor $\frac12$ in the definition of $\rho(\vec x)$ has
cancelled the factor 2 arising from $\text{Tr}(I)$.

Given this background information, we are ready to prove
Theorem~\ref{no-go}.  Suppose, toward a contradiction, that we have a
probability representation (Spekkens version) satisfying
determinateness and convex-linearity, for a two-dimensional \scr H.
In view of Proposition~\ref{translinear}, we know that
\[
\mu_{\rho(\vec x)}(\lambda)=\vec x\cdot\vec A(\lambda)+C(\lambda)
\]
and
\[
\xi_{E(m,\vec p)}= \vec p\cdot\vec B(\lambda)+m D(\lambda)+F(\lambda)
\]
for some nine functions $A_i(\lambda), B_i(\lambda), C(\lambda),
D(\lambda), F(\lambda)$ where the index $i$ ranges from 1 to 3.  (The
``translated'' part of ``translated-linear'' accounts for $C$ and
$F$.)

The definition of probability representation (Spekkens version) leads
to some simplifications.  $E(0,\vec0)$ is the zero operator, whose
associated $\xi$ function is required to be identically zero.  That
gives us $F(\lambda)=0$ for all $\lambda$, so we can simply omit $F$
from the formula for $\xi$.

Also, $E(1,\vec 0)$ is the identity operator, whose associated $\xi$
function is required to be identically 1.  That gives us
$D(\lambda)=1$ for all $\lambda$.  So we can simplify the $\xi$
formula above to read
\[
\xi_{E(m,\vec p)}= \vec p\cdot\vec B(\lambda)+m.
\]

Next, consider the requirement that
\[
\text{Tr}(\rho(\vec x)E(m,\vec p))=\int\xi_{E(m,\vec
  p)}\mu_{\rho(\vec x)}\,d\lambda.
\]
We already evaluated the trace on the left side of this equation at
the end of the preceding section.  The integral on the right side is
\[
\int[(\vec p\cdot\vec B(\lambda))(\vec x\cdot\vec A(\lambda))+
(\vec p\cdot\vec B(\lambda))C(\lambda)+
m(\vec x\cdot\vec A(\lambda))+mC(\lambda)]\,d\lambda.
\]
Comparing the trace and the integral, and equating coefficients of the
various monomials in $m$, $\vec p$, and $\vec x$, we find that
\begin{align}
\int B_i(\lambda)A_j(\lambda)\,d\lambda&=\delta_{i,j}, \label{2}\\
\int B_i(\lambda)C(\lambda)\,d\lambda&=0, \label{3}\\
\int A_i(\lambda)\,d\lambda&=0, \text{and} \label{4}\\
\int C(\lambda)\,d\lambda&=1.\label{5}
\end{align}

Next, we extract as much information as we can from the assumption
that all the functions $\mu_\rho$ and $\xi_E$ are nonnegative.

In the case of $\xi_E$, this means that, as long as $\nm{\vec p}\leq
m,1-m$ (so that $E(m,\vec p)$ is an effect), we must have $m+\vec
p\cdot\vec B(\lambda)\geq 0$ for all $\lambda$.  Temporarily consider
a fixed $\lambda$ and a fixed $m\in[0,\frac12]$. To get the most
information out of the inequality $m+\vec p\cdot\vec B(\lambda)\geq
0$, we choose the ``worst'' vector $\vec p$, i.e., we make $\vec
p\cdot\vec B(\lambda)$ as negative as possible, by choosing $\vec p$
in the opposite direction to $\vec B(\lambda)$ and with the largest
permitted magnitude, namely $m$.  That is, we take
\[
\vec p=-\frac m{\nm{\vec B(\lambda)}}\vec B(\lambda)
\]
so that our inequality becomes $0\leq m(1-\nm{\vec B(\lambda)})$, and
therefore
\[
\nm{\vec B(\lambda)}\leq1
\qquad\text{for all }\lambda.
\]
Repeating the exercise for $m\in[\frac12,1]$ gives no new information.

So we turn to the case of $\mu_{\rho(\vec x)}$, for which the
nonnegativity requirement reads
\[
\vec x\cdot\vec A(\lambda)+C(\lambda)\geq 0.
\]
For each fixed $\lambda$, we consider the ``worst" $\vec x$, namely a
vector $\vec x$ in the direction opposite to $\vec A(\lambda)$ and
with the maximum allowed magnitude, namely 1.  So we take
\[
\vec x=-\frac{\vec A(\lambda)}{\nm{\vec A(\lambda)}}
\]
and obtain the inequality $0\leq -\nm{\vec A(\lambda}+C(\lambda)$.
Thus, we have
\[
\nm{\vec A(\lambda)}\leq C(\lambda)\qquad\text{for all }\lambda.
\]
In particular, $C(\lambda)$ is everywhere nonnegative.

A trivial consequence of $\nm{\vec A(\lambda)}\leq C(\lambda)$ is that
$|A_1(\lambda)| \leq C(\lambda)$.  Similarly, a trivial consequence of
$\nm{\vec B(\lambda)}\leq1$ is $|B_1(\lambda)|\leq 1$.  Putting this
information into the $i=j=1$ case of equation~\eqref{2}, and also
using \eqref{5}, we find that
\[
1=\left|\int B_1(\lambda)A_1(\lambda)\,d\lambda\right|
\leq\int |B_1(\lambda)|\cdot|A_1(\lambda)|\,d\lambda
\leq\int 1\cdot C(\lambda)\,d\lambda=1.
\]
So both of the inequalities here must be equalities.  In particular,
$|B_1(\lambda)|=1$ for almost all $\lambda$ except where
$C(\lambda)=0$.

Similarly, we get that, for almost all $\lambda$ except where
$C(\lambda)=0$, we also have $|B_2(\lambda)|=|B_3(\lambda)|=1$ and
therefore $\nm{\vec B(\lambda)}=\sqrt3$.  Since we also know $\nm{\vec
  B(\lambda)}\leq 1$, we must conclude that $C(\lambda)=0$ almost
everywhere.  But that contradicts equation~\eqref{5}, and so the proof
of the no-go theorem is complete.
  \end{pf}

\end{document}